\documentclass[12pt,preprint]{aastex}
\usepackage{emulateapj5}

\begin{document}

\title{Formation of anomalous globular
clusters with metallicity spreads: A unified picture}

\author{Kenji Bekki} 
\affil{
ICRAR,
M468,
The University of Western Australia
35 Stirling Highway, Crawley
Western Australia, 6009, Australia
}

\and

\author{Takuji Tsujimoto}
\affil{
National Astronomical Observatory of Japan, Mitaka-shi, Tokyo 181-8588, Japan}

\begin{abstract}

Recent observations have revealed that at least 8 globular clusters
(GCs) in the Galaxy show internal abundance spreads in [Fe/H].
We investigate the origin of these `anomalous' GCs
using numerical simulations
of GCs in the dwarfs orbiting around the Galaxy and
chemical evolution
model of dwarfs hosting the GCs.
The principal results are as follows.
GCs formed in a host dwarf galaxy with a total mass
of $\sim 10^{10} {\rm M}_{\odot}$ can  merge to form a single nuclear GC
before the host is completely destroyed by the Galaxy,
if they are massive ($> 3 \times 10^5 {\rm M}_{\odot}$)
and if they are formed in
the inner region ($R<400$ pc).
The GC merger remnants can capture field stars during
its spiral-in to nuclear regions.
If two GCs are formed from star formation events  separated by $\sim 300$ Myr
in their host dwarf,
then the new GC formed from GC merging can have
[Fe/H] spread of 0.2 dex and [Ba/Fe] spread of 0.3 dex.
GCs formed from GC merging can show variety of internal abundance spreads
depending on the details of their hosts' chemical evolution.
We suggest that
anomalous GCs were formed from GC merging that occurred before the destruction
of GC host dwarfs yet after self-enrichment processes responsible
for the observed anti-correlations between chemical abundances
of light elements.
We also suggest that the observed no/little dependence of [Eu/Fe]
on [Fe/H] in the Galactic GC  M22 is  evidence of massive dwarf galaxies
hosting these anomalous GCs.

\end{abstract}

\keywords{
globular cluster: general --
galaxies: star clusters: general --
galaxies: stellar content --
stars:formation
}

\section{Introduction}

Recent photometric and spectroscopic observations
of GCs have established  
the presence of multiple stellar populations in 
most of the investigated Galactic GCs (e.g., Gratton et al. 2012 for
a recent review). Strong evidence for the multiple stellar populations
includes, for example, the multiple main-sequence turn-offs
and triple main sequence in $\omega$ Cen and NGC 2808
(e.g., Bedin et al. 2004; Piotto et al. 2005; 2007),
ubiquitous anti-correlations between light elements 
in most Galactic GCs (e.g., Carretta et at. 2009),
and internal star-to-star variations in $s$-process elements
in some massive GCs (e.g., Yong et al. 2014; Marino et al. 2015; M15).
These observations have raised a number of fundamental questions
regarding the GC formation and evolution: (i) whether or not
the presence of
multiple stellar populations means the  multiple generations of 
stars (e.g., Bastian et al 2013; Renzini et al. 2015),
(ii) how  the observed C-N and O-Na anti-correlations
can be explained by theoretical models based on the self-enrichment
by asymptotic giant branch stars (AGB) or fast-rotating massive
stars (FRMS) in the early evolution of GCs (e.g., Decressin
et al. 2007; D'Ercole et al. 2008; D08),
and (iii) how multi-generation of stars can be successively formed 
in GCs
(e.g., D08; Bekki 2011; B11).

One of the key unresolved problems related to  multiple stellar 
populations of GCs is the origin of `anomalous GC' with internal
metallicity spreads ($\Delta$[Fe/H]$>0.05$ dex) among GC stars
(See Table 10 in  M15 for a nice summary
of chemical properties for the 8 anomalous GCs).
Each of these 8 GCs is observed to show internal spreads  
in chemical abundances of stars other than [Fe/H],
such as C+N+O and $s$-process elements.
For example, Marino et al. (2011) found that 
(i) M22 has at least two groups of stars with the [Fe/H] difference
of $\sim 0.15$ dex among the two groups,
(ii)  Fe-rich  group also shows a higher abundance of 
$s$-process element,
and (iii) C-N and Na-O anti-correlation can be seen in
each of the two groups.
Yong et al. (2014) have recently revealed the presence of three
distinct populations with [Fe/H] of $\approx -1.7$, $-1.5$ and $-1.0$
and internal abundance spreads in $s$-process elements in the
Galactic GC M2
and thus suggested that the formation history of M2
can be quite similar to other anomalous GCs such as $\omega$ Cen
(See Lardo et al. 2013 for similar results on
the abundance spread in M2).
Internal abundance spreads in $s$-process elements, $p$-capture
elements, and  C+N+O are observed in NGC 1851 (e.g.,
Yong \& Gundhahl (2008), Carretta et al. (2010),
Villanova et al. (2010), and Youg
et al. (2014).

Johnson et al. (2015) have recently investigated 
chemical properties of stars in 
NGC 6273 (M19), which is not listed as an anomalous GC in M15,
and found $\Delta$[Fe/H] of $0.5$ dex among
the GC stars, a significant [La/Fe] enhancement in more metal-rich
stars, and a possible third metal-rich population.
Lee et al. (2009) investigated the $hk$ index in RGB sequences of 37
Galactic GCs and found that more than  half of them show discrete or broad
distributions of the $hk$ index. If $\Delta$[Ca/H] among GC stars
is responsible
for the observed broad $hk$ index distribution, then
self-enrichment by supernovae needs to occur in most of  GCs
at their formation phases.
The number of anomalous GCs
appears to be currently increasing whereas the latest observations
of massive GCs
(e.g., NGC 2808 and 47 Tuc) does not show large  [Fe/H]
spreads (e.g., Carreta 2015; Milone et al. 2015; Marino et al. 2016).
These imply that
chemical evolution processes in forming GCs is more complicated
than previous theoretical models of GC formation predicted 
(e.g., D'Ercole et al. 2008; B11).

In spite of these recent observational developments,
theoretical studies of GC formation have not yet discussed
the origin of the observed unique chemical and dynamical properties
of anomalous GCs extensively.
Bekki \& Yong (2012, BY12) discussed the origin of the stellar halo
around NGC 1851, which is one of anomalous GCs in the Galaxy,
in the context of GC formation from 
dwarf galaxy with stellar galactic nuclei. 
D'Antona et al. (2016) have recently constructed a new model
in which the detailed
star formation history and chemical enrichment
process by Type II supernovae and AGB stars
are derived  for explaining the observed chemical abundances
of the major five stellar populations in NGC 2808.
These recent theoretical studies, however,
discussed either only the dynamical process of GC formation
or only the chemical evolution process of forming GC.
Theoretical models
that can predict both various chemical abundances
(e.g., $\alpha-$, $s-$, and $r-$process elements)
and dynamical processes of GC formation and evolution
(e.g., GC merging and tidal stripping of GC stars)
are thus required for better understating the origin of the 
anomalous GCs.

The purpose of this paper is thus to investigate
{\it both} dynamical evolution of GCs in host dwarf galaxies
and chemical evolution of the GC hosts
in order to make a more comprehensive understanding
of the origin of the Galactic anomalous GCs.
We adopt a new dynamical model in which both GCs and their host dwarfs
are represented by  N-body particles in a fully self-consistent
manner so that dynamical friction of GCs against field star and dark
matter,  GC merging, and tidal stripping of GC stars can be 
investigated. 
Therefore, this model is significantly better than our previous model used
in BY12.
Furthermore, using 
one-zone chemical evolution models for dwarf galaxies
(e.g., Tsujimoto 2011, T11; Tsujimoto \& Bekki 2013;
Tsujimoto \& Shigeyama 2014),
the present study investigates
the time evolution of various chemical
abundances, such as [Fe/H], [Ba/Fe], and [Eu/Fe] and their dependences
on the star formation histories and the initial stellar mass function
(IMF). The results can be interpreted in the context of the chemical
abundances of anomalous GCs.

The plan of the paper is as follows.
We describe the new dynamical evolution models of dwarf galaxies
with two GCs orbiting around the Galaxy
in \S 2.
We present the numerical results on the possibility of GC merging
in dwarfs and its dependence on the physical properties of GCs
and dwarfs in \S 3.
We mainly discuss whether the large spreads in
[Fe/H] and $s$-process elements observed in the anomalous
GCs can be possible in new GCs
formed from  GC merging using one-zone chemical evolution models in \S 4. 
We  provide important implications of the present results in terms of
the origin of multiple stellar populations in GCs in this section
in \S 5.
We summarize our  conclusions in \S 6.
It should be noted that we do not numerically investigate the formation of
GCs from gas clouds within high-$z$ dwarfs in the present paper:
We have recently investigated how GCs can be formed from massive
gas clumps formed in gas-rich dwarf galaxies using hydrodynamical
simulations of dwarfs (Bekki 2016).

\section{The model for GC merging}

\subsection{Two-fold investigation}

We adopt the GC merging scenario for the formation of
anomalous GCs, which is  illustrated in Figure 1.
The present study consists of (i) dynamical simulations of GC mergers
in GC-host dwarf galaxies and (ii) chemical evolution study of the GC hosts.
We first investigates the dynamical evolution of two GCs in massive
dwarf galaxies using our own original simulation code that can be run
on GPU clusters (Bekki 2013). In this first investigation,
the probability of GC merging and the merging timescale of two GCs
in dwarfs are discussed in detail. Although this dynamical simulations is 
similar to BY12, it is more sophisticated than BY12 in that (i)
two GCs are represented by N-body particles in a self-consistent manner
and (ii) GCs can be represented by FG+SG systems or  by FG-only systems.
Therefore, the present study can make more precise predictions on
the dynamical fate of GCs in massive dwarfs.

We also investigate the chemical evolution of dwarfs, in particular,
the evolution of [Fe/H], [$s$-process elements/Fe], and 
[$r$-process elements/Fe] in dwarfs, using one-zone chemical evolution models
developed by T11.
The present study considers that chemical abundances of anomalous GCs
can reflect the chemical evolution of their host dwarfs
and thereby investigates the possible spreads in the above chemical
abundances  within GCs. We do not discuss self-enrichment processes 
in the formation of SG stars from ejecta from FG stars, because
they have been discussed by many authors (e.g., Fenner et al. 2004;
Bekki et al. 2007;  D'Ercole et al. 2010).
Since the numerical methods for the dynamical evolution of dwarfs
with GCs in the Galaxy
were already given in BY12,  we briefly describe them below.
For clarity,
we describe the one-zone chemical evolution models originally
developed by T11
and the results of the models  in a separate section, \S 4.

\subsection{Dwarf galaxy}

A GC-host dwarf galaxy is  
assumed to consist of a dark matter halo
and a stellar disk without nucleus.
Both the dark matter halo and the stellar disk of the dwarf
are represented by collisionless N-body particles.
Gas dynamics, star formation, chemical evolution,
and dust formation 
are not included at all in the present study (i.e.,
the simulations are purely collisionless ones).
The dark matter halo with 
with the total mass of $M_{\rm h}$ is represented by
the `NFW' one (Navarro et al. 1996)
with a central cusp predicted by the Cold Dark Matter (CDM)  model:
\begin{equation}
{\rho}(r)=\frac{\rho_{0}}{(r/r_{\rm s})(1+r/r_{\rm s})^2},
\end{equation}
where $r$,  $\rho_{0}$,  and $r_{\rm s}$ are the distance from the center
of the cluster, the central density, and the scale-length of the dark halo,
respectively.
The virial radius ($r_{\rm vir}$),  the scale radius ($r_{\rm s}$),
and the `$c$' parameter (=$r_{\rm vir}/r_{\rm s}$)
are chosen such that the values
are consistent with recent cosmological simulations 
for the adopted $M_{\rm h}$
(Neto et al. 2007).

For comparison, we 
also investigate the models with   `cored
dark matter' halos (Salucci \& Burkert 2000; 'SB' profile),
the density profile of which is described as follows:
\begin{equation}
{\rho}(r)=\frac{\rho_{\rm 0,dm}}{(r+a_{\rm dm})(r^2+{a_{\rm dm}}^2)},
\end{equation}
where $\rho_{\rm 0,dm}$ and $a_{\rm dm}$ are the central dark matter
density multiplied by $a_{\rm dm}^3$
and the core (scale) radius, respectively.
Bekki et al. (2003) already demonstrated that
dwarf galaxies with SB profiles are more susceptible to tidal
destruction by larger and more massive halos 
than those with NFW profiles
and discussed the transformation of nucleated dwarfs into naked
nuclei.
Therefore, we do not discuss much about dwarf destruction processes
for dwarfs with SB profiles and GC evolution within them. We 
show the results for only one model with $M_{\rm h}=10^{10} M_{\odot}$,
$a_{\rm dm}=0.92$ kpc, and $r_{\rm vir}=9.2$ kpc.

The dwarf is assumed to be as a  bulge-less disk galaxy
with the total stellar mass of $M_{\rm s}$ and the size of $R_{\rm s}$.
The radial ($R$) and vertical ($Z$) density profiles of the stellar disk are
assumed to be proportional to $\exp (-R/R_{0}) $ with scale
length $R_{0} = 0.2R_{\rm s}$ and to ${\rm sech}^2 (Z/Z_{0})$ with scale
length $Z_{0} = 0.04R_{\rm s}$ , respectively.
In addition to the
rotational velocity caused by the gravitational field of disk
and dark halo components, the initial radial and azimuthal
velocity dispersions are assigned to the disc component according to
the epicyclic theory with Toomre's parameter $Q$ = 1.5.  The
vertical velocity dispersion at a given radius is set to be 0.5
times as large as the radial velocity dispersion at that point.
The stellar disk is assumed to have no stellar nucleus initially,
which is different from the initial disk model adopted by BY12.

\subsection{GCs}

A dwarf galaxy is assumed to have  two GCs located in different
regions within the dwarf. Two GCs are denoted GC1 and GC2 for convenience,
and they are represented by collisionless particles.
GC1 and GC2 have different total mass 
($m_{\rm gc,1}$ and $m_{\rm gc, 2}$, 
respectively)
and distances ($R_{\rm gc, 1}$ and $R_{\rm gc, 2}$, respectively). 
For convenience,  the larger GC is referred to as GC1
in the present study.
Recent theoretical
studies of GC formation have shown that 
new stars can be formed from gas ejecta
from AGB stars of GCs (D08, B11).
The new stars and the existing stars are often referred to
`SG' and `FG' stars, respectively, and we adopt these abbreviations
in the present paper.
The SG systems are demonstrated by these previous works
to be  more compact
than the FG ones, because SG stars can form in the deep potential
wells of FG stellar systems.
Following these  results, we mainly investigate
`two-component' GC models in which a GC is composed of FG and SG 
systems with different masses (e.g., $m_{\rm FG, 1}$ and $m_{\rm SG, 1}$
for GC 1)
and sizes ($r_{\rm FG, 1}$ and $r_{\rm SG, 1}$ for GC1). 
For comparison,
we also investigate `one-component' GC models in which a GC is composed
only of a FG system.

Each sub-population (FG or SG) in a GC is assumed to have the radial
density profiles of stars described by a Plummer model  with  
the scale length ($a$) being 0.2 times the size of the stellar system
(e.g., $a_{\rm FG, 1}=0.2 r_{\rm FG, 1}$). Each cluster has 3D velocities
that are the same as those of its nearest star in the stellar disk
of the GC-host dwarf. 
The mass-ratio of two GCs ($m_2=M_{\rm gc, 2}/M_{\rm gc,1}$,
where $M_{\rm gc, 1}$ is always larger than or equal to $M_{\rm gc,2}$)
is one of the most important parameters
that can determine whether the GCs can merge with each other
before the complete destruction of their host dwarfs.
We mainly investigate the models in which $m_{\rm FG, 1}=m_{\rm FG, 2}$
(i.e.,  major GC merging), because BY12 have shown
that major merging is  more likely to occur than minor merging with 
$m_{\rm FG, 1}/m_{\rm FG, 2} < 0.3$ within dwarfs.

Vesperini et al. (2013) investigated the dynamical evolution
of GCs with FG and SG stars that have initially different
spatial distributions using N-body simulations. They found
that complete spatial mixing of the two stellar 
populations due to two-body relaxation  is possibly
only in the advanced stage of dynamical evolution.
We do not discuss the two-body relaxation effects on the evolution
of merged GCs with initially two populations
in the present study, though it is a quite interesting issue.
Recently Ganagnin et al. (2016) have extensively investigated
the structure and kinematics of the remnant of GC mergers for 
a wide range of densities and mass-ratios of merging GCs.
Although these results are quite useful for understanding the origin
of rotation in GCs (e.g., Anderson \& King 2003; Bekki 2010a),
we do not conduct such an extensive study of structural and
kinematical properties of merged GCs, because our main interest
is the probability of GC merging.

\subsection{Dwarf orbit and the gravitational potential
of the Galaxy}

We mainly investigated two models, `the present' and `young' Galaxy
(Milky Way; MW) models to discuss the GC merging and stripping in early and late
phases of the Galaxy evolution.
The model parameters for halo, bulge, and disk components
of the Galaxy are different between the present and young MW models.
The Galaxy in present MW models is assumed to have a 
{\it fixed} three-component gravitational potential
and the orbit of the  dwarf   can be determined
by the details of the potential for given its initial 3D positions,
($x$, $y$, $z$),
and 3D velocities, ($v_{\rm x}$,$v_{\rm y}$,$v_{\rm z}$),
with respect to the Galactic center.
The following  logarithmic dark matter halo potential
is adopted for the Galaxy,
\begin{equation}
{\Phi}_{\rm halo}=v_{\rm halo}^2 \ln (r^2+d^2),
\end{equation}
where
$d$ = 12 kpc, $v_{\rm halo}$ = 131.5 km ${\rm s}^{-1}$ and
$r$ is the distance from the center of the Galaxy.
The gravitational potential of the Galactic disk is represented by
a Miyamoto-Nagai (1975) potential;
\begin{equation}
{\Phi}_{\rm disk}=-\frac{GM_{\rm disk}}{\sqrt{R^2 +{(a+\sqrt{z^2+b^2})}^2}},
\end{equation}
where $M_{\rm disk}$ = 1.0 $\times$ $10^{11}$ $M_{\odot}$,
and $a$ = 6.5 kpc, $b$ = 0.26 kpc,
and $R=\sqrt{x^2+y^2}$.
The following  spherical Hernquist (1990) model is adopted for
the potential of the Galactic bulge;
\begin{equation}
{\Phi}_{\rm bulge}=-\frac{GM_{\rm bulge}}{r+c},
\end{equation}
where $M_{\rm bulge}$ =  3.4
$\times$ $10^{10}$ $M_{\odot}$,
and $c$ = 0.7 kpc. This reasonable set of parameters gives a realistic
rotation curve for the Galaxy with a maximum
rotation speed of 224 km ${\rm s}^{-1}$ at $R=8.5$ kpc.

For the young MW models,
the above three potential profiles with different parameters for
the total masses of the three components  are adopted: 
$M_{\rm disk}$ = 1.0 $\times$ $10^{10}$ $M_{\odot}$,
$M_{\rm bulge}$ = 3.4
$\times$ $10^{9}$ $M_{\odot}$,
and 
$v_{\rm halo}$ = 93.0 km ${\rm s}^{-1}$. This combination of 
parameters is chosen to mimic the young MW 
(at the  time of major  GC formation) with a possibly
 much lower mass
and thus a  lower maximum circular velocity.
The epoch of dwarf accretion onto the Galaxy can be quite 
different between different dwarfs with GCs
and the Galactic potential should be time-evolving owing to
hierarchical mass accretion. By using the two extreme cases
of the Galactic gravitational potential,
we try to understand how the Galactic tidal field can influence
GC merging during dwarf
destruction.

The 3D position of the Galactic center  is  fixed
at ($x$,$y$,$z$) = (0,0,0) whereas the initial 3D position and velocity
of a dwarf galaxy are
free parameters that can determine the orbital evolution of the dwarf
around the Galaxy.
The three basic parameters are 
(i) the initial distance of the dwarf from the Galactic center
($R_{\rm i}$),
(ii) the initial velocity $f_{\rm v}v_{\rm c}$,
where $v_{\rm c}$ is the circular velocity at $R_{\rm i}$
and $f_{\rm v}$ is a parameter that controls the orbital eccentricity
and ranges from 0.1 (highly eccentric orbit) to  1 (circular orbit).
and (iii) the inclination angle between the Galactic disk plane 
(=the $x$-$z$ plane)
and the orbital plane of the dwarf ($\theta$).
The initial $x$-component of the velocity of the dwarf
is set to be 0, which means that $R_{\rm i}$ corresponds
to the apocenter distance of the dwarf's orbit,
because $f_{\rm v}$ is always less than 1.
We mainly investigate the model (`standard orbit', defined
as `O1') in which
$R_{\rm i}$=17.5 kpc,  
$f_{\rm v}=0.5$, and $\theta=45^{\circ}$,
because dwarf galaxies can be destroyed almost completely
within $\sim 3$ Gyr in this model.
We also investigate models with different $R_{\rm i}$, $f_{\rm v}$,
and $\theta$ in order to investigate how the details of the orbital
evolution of dwarfs can influence the merging processes of GCs.
The initial angle between the stellar disk of a dwarf
and the orbital plane of the dwarf is 60 degrees for all models
in the present study.

\subsection{Parameter study}

Although we investigated numerous models (56 models), we present the results
of 19 representative models,  for which the model parameters
are described in Table 1.
The model parameters for 6 dwarf galaxy models
investigated
in the present study (D1-D6)
and the adopted orbits of the dwarfs (O1-O4)
are summarized in Table 2 and 3,
respectively.
Two GCs are assumed to exist initially in a simulated dwarf,
because we think that the time delay between the formation epochs
does not influence their evolution within the dwarf.
 We mainly show the results of the 
fiducial model (M1) with  the dwarf galaxy model D1
($M_{\rm h}=10^{10}$ ${\rm M}_{\odot}$)
$m_{\rm FG, 1}=10^6 M_{\odot}$,
$a_{\rm FG, 1}=10$ pc, 
$m_2=1$ (the mass-ratio of two GCs),  $R_{\rm gc, 1}=100$ pc,
$R_{\rm gc, 2}=200$ pc,
and the standard orbit model O1,
because it clearly shows 
a typical behavior of GC merging 
in a massive dwarf galaxy being destroyed by the Galaxy.
We mainly focus on the possibility and timescale of GC merging
in dwarf galaxies and their dependency on model parameters
such as $M_{\rm h}$ and $m_{\rm FG. 1}$. 
The timescale of GC merging ($t_{\rm merge}$) for each model is given in Table 1.
We define $t_{\rm merge}$ as the time when (i) the separation of two GCs becomes
less than 10pc and (ii) the separation becomes the minimum for the first time.
The above condition (i) is necessary, because some models do not show GC merging.
We also investigate
the fraction of SG stars in the halo regions around merged
GCs in order to discuss the observed physical properties
of stellar halos around GCs (e.g., Olszewski et al. 2009).
GC merging process can be influenced by the presence of stellar galactic nuclei,
however, we do not discuss this possible important effect in the present study.
Appendix A discusses briefly this issues using a few models with stellar galactic
nuclei.

We investigate the evolution of dwarfs with GCs
over 2.82 Gyr and 5.64 Gyr for the models
with smaller $R_{\rm i}$ (=8.5 and 17.5 kpc)
larger initial $R_{\rm i}$ (35 kpc), respectively.
We run a low-mass dwarf model ($M_{\rm h} = 10^8 M_{\odot}$)
without GCs (M19) in order to
show the orbit of a dwarf that is not influenced by GCs 
(e.g., dynamical friction) and
stripped field stars for a given gravitational
potential: a dwarf's orbit defined
by the time evolution of the position of the central star
can be slightly influenced by
massive GCs and stripped stars.
GCs can continue to lose a  significant fraction ($\sim 60$\%)
of their stars through two-body relaxation and 
tidal stripping by the Galactic tidal field 
over the time scale of $\sim 10$ Gyr (e.g., Vesperini 1997; Vesperini et al.
2011). These long-term effects are not properly included, because
we run simulations for only $3-6$ Gyr. Accordingly, the final masses
of GCs could be larger than the observed ones in the present study.
For example, if the present-day mass of M22 is $2.9 \times 10^5 M_{\odot}$
(e.g., Marks \& Kroupa 2010), then its  mass before
losing stars through the above long-term dynamical  effects should be
at least $4.8 \times 10^5 M_{\odot}$.  The initial mass of M22 at its birth
could be even
 by a factor of $5-10$ larger than this $4.8 \times 10^5 M_{\odot}$,
because  FG stars can be lost during the early evolution
of GCs (e.g., D'Ercole et al. 2008).

\subsection{Simulation set up}

We use a newly revised, more sophisticated version of our original  simulation
code adopted in BY12 so that
we can investigate not only  global dynamical evolution
of dwarf galaxies influenced by the Galactic tidal field
but also  the  orbital evolution of GCs within them.
The time integration of the equation of motion
is performed by using 2nd-order
leap-flog method with the maximum time
step interval being $1.41 \times  10^6$ yr
for particles in a dwarf galaxy and $2.82 \times 10^4$ yr
for particles representing GCs.
The adopted individual time step scheme
with rather small time step interval for GC stars enables
the present study to investigate GC merging more properly
than BY12,
because such scheme was not introduced by BY12.

The total number of particles
for dark matter, stellar disk,
GC  of a dwarf galaxy
is 500,000, 500,000,  and 60,000, respectively.
The FG and SG systems are represented by 20000 and 10000 collisionless
particles in all models.
Only  a limited amount of computations time is allocated
for the research project on which the present study is based
and we need to run at least $\sim 60$ models
with different model parameters  in the present investigation.
We therefore consider that
the total number particle of $\sim 10^6$ 
is  reasonable (not numerically costly) that we can adopt for each model.
The gravitational softening lengths ($\epsilon$)  
for  dark matter halo,
stellar disk,
FG system,
and SG system (of a GC)
are denoted as ${\epsilon}_{\rm dm}$,
${\epsilon}_{\rm s}$, 
${\epsilon}_{\rm FG}$, 
and ${\epsilon}_{\rm SG}$, 
respectively.
We determine $\epsilon$ for each of these 
components based on the half-number radius of
the particles.
We consider that
when two different components interact gravitationally,
the mean softening length for the two components
is applied for the gravitational calculation.
For example, $\epsilon = ({\epsilon}_{\rm dm}+{\epsilon}_{\rm s})/2$
is used for gravitational interaction between dark matter
particles and stellar ones  in a dwarf.
In the fiducial model,
${\epsilon}_{\rm dm}$, ${\epsilon}_{\rm s}$
${\epsilon}_{\rm FG}$, and ${\epsilon}_{\rm SG}$
are set to be 104\,pc,  10.8\,pc, 0.93\,pc,
and 0.37 pc,
respectively,
in the fiducial model (M1).
In the following, $T$ in a simulation
represents the time that has elapsed since the simulation
started.

\section{Results: GC merging}

\subsection{The fiducial model}

Figure 3  shows how the spatial distributions
of field stars of a dwarf disk galaxy and stars
from FG and SG systems in two GCs evolve with time during the 
accretion of the dwarf onto the  Galaxy
in the fiducial model M1. This model
corresponds to a later accretion of a dwarf onto the present Galaxy
so that the strong tidal field of the Galaxy can rapidly  destroy
the dwarf within $\sim 3$ Gyr.
A significant fraction
of FG stars can be efficiently stripped from GC1 and GC2
by the dwarf's tidal field while the two GCs are orbiting around the
host dwarf ($T=0.56$ Gyr). 
These stripped FG stars spread over the entire disk region
of the dwarf ($T=1.12$ Gyr) and then are stripped from the dwarf while
the stellar disk component
of the dwarf is being destroyed by the Galactic tidal field 
($T=1.68$ and 2.24 Gyr).
After the complete destruction  of the dwarf
by the Galactic tidal field, the new GC (or naked nucleus) can continue
to lose their FG and SG stars with FG stars being more efficiently
stripped by the tidal field. Clearly a tidal stream
connecting with the halo of the new GC can be seen at $T=2.24$ Gyr,
whereas there is no such connection at $T=2.82$ Gyr. 
The final total mass of GC stars 
at $T=2.82$ Gyr
is $6.2 \times 10^5 M_{\odot}$ 
for $R \le 10$ pc
and $1.2 \times 10^6 M_{\odot}$
for $R \le 50$ pc.

Since the dynamical friction of GCs against
disk field stars of the dwarf is quite effective owing to the lower velocity
dispersion of field stars (Bekki 2010b),
the two GCs can spiral into the central
region of the dwarf and finally merge with each other.
As shown in Figure 4, the two GCs can rather quickly  merge to form
a new GCs (a single nuclear GC)
at $T\sim 0.32$ Gyr. The SG systems do not lose a significant 
fraction of stars during major GC merging
so that the new SG system can be still more compact.
A significant fraction
of the FG stars can be stripped to form a stellar halo during GC merging.
The final remnant of GC merging can have diffuse FG and compact SG
system,  and the remnant can still have a nested structure 
($T=1.12$ Gyr).
The stellar halo around the new GC is quite extended within its host dwarf
and the halo is dominated by FG stars
at $T=0.56$ Gyr when the new GC can be identified as an off-center nucleus.
Thus, this fiducial model demonstrates how two massive GCs 
with FG and SG stars can be transformed
into one new GC with four distinct stellar populations
(i.e., original  FG and SG stars from GC1 and GC 2)  during the destruction
of their host galaxies.

Figure 5 shows how the mass fractions of disk field stars of the dwarf
($F_{\rm Field}$)
depends on the distance ($R$) from the center of the remnant of GC merging.
The mass fraction
is not so large ($<3$\%) in the halo region
of the new GC ($R<50$ pc; original radius of the FG system)
at $T=2.82$ Gyr, which means that
the field stars can not be tidally
captured by the merged GC so efficiently in this model.
These stars that are gravitationally trapped within the merged GC
(i.e., those with their velocities less than the escape velocity of the GC)
are mostly from the inner regions of the dwarf.
Both dwarf galaxies and the Galaxy
are observed to have  negative metallicity gradient 
(e.g., Andrievsky et al. 2004; Tolstoy et al. 2004
de Boer et al. 2012),
though these dwarfs with examined metallicity gradients
are dwarf spheroids rather than dwarf disks (irregulars).
Hidalgo-Gamez et al. (2010) revealed
negative abundance gradients in [Fe/H] and oxygen with $\sim -0.2$ dex kpc$^{-1}$
in dwarf spirals, though their paper is yet to be published.
Given these observations,
the trapped stars can  
be possibly more metal-rich in some dwarf disk galaxies.
The derived small fraction of field stars around the new GC
is consistent with the previous results by BY12.

A significant contribution of field stars ($\sim 20$\%)
to the halo region of the GC
can be seen only in the very
outer part of the new GC ($R\sim 100$ pc), though not all of these
outer halos stars are gravitationally bound to the new GC.
The mass fraction of SG stars within the central 10pc of the new GC 
($\sim 60$\%) is consistent with the observed fraction of SG stars
(e.g., Carretta et al. 2009).
A much  smaller fraction of stars can be stripped from
the SG system of the new GCs so that the halo of the new cluster
can not be dominated by the stripped SG stars (at most $\sim 20$\%). 
The presence of SG stars in the simulated GC is not so consistent with
recent observations which have not yet discovered the presence of SG stars
in the halos of examined GCs (e.g., NGC 1851; Marino et al. 2015).
However, the lack of SG stars in these GCs could be simple due to
the expected small number of SG stars in the halo regions.

Figure 5 shows the separate rotation curves ($V(R)$) of the FG and SG systems
of the new GC at $T=2.82$ Gyr.
The $V(R)$ profiles  along the $y$-axis are derived
using  the $x$-components of line-of-sight
velocities for stars within 50pc from
the GC center are used at each radial bin.
Both FG and SG stars are being stripped at  $T=2.82$ Gyr,
and therefore the stars connecting to the tidal streams introduce a larger
velocity dispersion.
The global rotation of the GC can be seen slightly more clearly in the SG system
than in the FG, though the rotation amplitude is small ($\sim 1.5$ km s$^{-1}$)
for FG and SG stars. There is little  rotation in the central 10 pc both for
the FG and SG stellar systems. Although the amplitude of rotation of merged GCs
depends on the viewing angles and the orbits of merging GCs,
it is quite small (at most 5 km s$^{-1}$ in different projections).

\subsection{Parameter dependence}

It is found that the  total masses of GCs,  the initial positions of
GCs within their host dwarfs,  and the physical properties of GC hosts,
and the orbits of GC hosts can determine the probability of GC merging.
These results, some of which are essentially the same
as those reported in BY12, are summarized as follows.  \\

(1) Figure 6 shows that if both GCs have $r_{\rm gc} \le  600$ pc,
then the two GCs can merge with each other well before
the complete destruction of their host in the dwarf model D1
(See M2 and M3).
Such GC merging within their host can not occur in the model M4
in which one of the two GCs is initially located in the outer part
of the host ($R= 800$ pc).
This results can be understood as follows.
The timescale of a GC to spiral in the nuclear regions of its host dwarf
due to dynamical friction is longer for larger $r_{\rm gc}$,
and GCs with larger $r_{\rm gc}$ are more likely to be stripped
from their host dwarf during the tidal interaction between the dwarf
and the Galaxy. Therefore, only GCs initially in the central
regions are expected to merge with one another before 
they are tidally stripped from their hosts. \\

(2) GC merging is unlikely  within $\sim 2$ Gyr (=timescale of dwarf
destruction; $t_{\rm dest}$) 
in the model M5 with $m_{\rm FG, 1}=3 \times 10^5 M_{\odot}$  
and in the model M6
with $m_{\rm FG, 1}=10^5 M_{\odot}$, as shown in Figure 6.
These results suggest that there could be a threshold GC mass above which
GC merging is possible in dwarfs. These results also imply that
more massive GCs (e.g., M22)
are more likely to become GCs with metallicity spreads
though GC merging,
because GCs formed in different local regions within their dwarfs
at different times
are likely to have different metallicities. \\

(3) If the mass-ratio of two GCs is low ($m_2=0.1$), then
GC merging within $t_{\rm dest}$ is not possible even for 
$m_{\rm FG, 1}=10^6 M_{\odot}$, as shown for the model M7 in Figure 6.
This is because the timescale of dynamical friction for the smaller
GC is quite long: the two GCs can not become close enough to merge in
the central region of their host dwarf.
This result implies that the mass ratios of two populations with
different metallicities in a GC does not differ so much, if
GCs with metallicity spreads are formed from GC merging.
The mass-ratios of two populations with different [Fe/H] in anomalous
GCs such as NGC 1851 and M22 have not been precisely determined
or clearly  documented yet (e.g., Marino et al. 2011).
The present simulations predict that  the mass-ratios of two populations
with different [Fe/H] should be comparable in anomalous GCs.  \\

(4) If GC-host dwarf galaxies are as massive as 
$M_{\rm h}= 3 \times 10^{10} M_{\odot}$ (corresponding to the Magellanic
Clouds), then GC merging becomes less likely even for massive GCs with
$m_{\rm FG, 1}=10^6 M_{\odot}$. 
This is mainly because dynamical friction time scale is longer in more massive
dwarfs for a given GC mass:
if the relative velocity between
the GC and the field stars is larger
(as is the case for more massive dwarfs), then the dynamical friction time scale
is longer. 
The model M11 with the massive
dwarf model D4 ($M_{\rm h}= 3 \times 10^{10} M_{\odot}$) do not 
show GC merging if two GCs are located at $R=100 \sim 200$ pc.
This result suggests that formation of anomalous GCs through
GC merging is possible in dwarfs that are not too massive. 
It is confirmed that major merging of two GCs with 
with $m_{\rm FG, 1}=3 \times 10^6 M_{\odot}$ is possible
in very massive dwarfs
with $M_{\rm h}= 3 \times 10^{10} M_{\odot}$,
though the GC merger remnant is quite massive (like $\omega$ Cen).
 This indicates
that the mass-ratios of GCs to dark matter halos ($f_{\rm m, gc}$)  is a key
parameter for GC merging. \\

(5) It is confirmed that the probability of GC merging does not 
depend on whether or not GCs are composed only of FG stars 
or both of FG and SG
stars: The model M12 with only FG stars demonstrates GC merging within
less than 1 Gyr. Also, the models M14, M15, and M16 show that GC merging
is possible in the young Galaxy model, which strongly suggests that
the formation of anomalous GCs through GC merging within dwarf is possible
in the early formation history of the Galaxy.
Although the timescale of GC merging becomes longer for larger
initial distances of dwarfs (e.g., M18 for O2 orbit),
GC merging process itself is not strongly influenced by
the initial distances of GC host dwarfs.
The dark matter density profiles of dwarfs do not influence
GC merging so greatly,  if GC masses are  large  enough
($m_{\rm FG, 1}=10^6 M_{\odot}$), as shown in the model M17. \\

(6) Figure 7 shows that the mass fractions of field stars in the halos
of merged GCs does not depend so much on model parameters 
($F_{\rm field} \sim 0.1$). The presence of field stars within 10-20 pc
of merged GCs can be seen only if the initial masses of individual GCs to
their hosts ($f_{\rm m, gc}$) are higher ($f_{\rm m, gc} \sim 0.0001$).
Since the field stars finally within merged GCs are mostly from the inner
region of dwarfs, field stars within
the GCs (i.e., anomalous GCs) 
are likely to be observed as anomalous metal-rich stars. 
GC1 in the model M4 in which GC merging is not possible have no field stars
in the outer part, because the GC is tidally stripped from the host dwarf
so that it can have little time to tidally capture field stars.
This results suggests that GC merging is essential to host field stars
in the halo regions of GCs.  The merged GC in the model M9 has
a large mass ratio of field stars to FG stars in its halo, because
the destruction of its host dwarf has just been complete at $T=2.8$ Gyr
and therefore disk field stars can still remain in the surrounding region
of the GC.
This result means that anomalous GCs just after their `release'
from the host dwarfs can have large halos dominated by field stars. \\

(7) The dwarf galaxy in the model M13 without GCs can not be completely
destroyed by the Galaxy within $\sim 3$ Gyr, which is in  striking 
contrast with the fiducial model with GCs
which shows complete tidal destruction
of the dwarf galaxy within 2 Gyr. This result implies that
spiral-in of GCs can lower the central density of the dark matter
halo of its host dwarf so that the dwarf becomes more susceptible
to tidal destruction after GC merging. 
This role of GC merging in the destruction process
of GC host dwarf  is a preliminary result and will need to be re-investigated
 more extensively in our future studies. \\

(8) The present results do not depend strongly on the models of the Galaxy,
the central densities of dark matter halos of dwarfs, and the orbits
of th dwarfs. For example, as shown in Table 1 for M14 - M16,
merging of more massive GCs in dwarfs is possible for the young
Galaxy model. Also, GC merging is possible in the cored dark matter
model, M17, and in M18 in which the apocenter of the dwarf orbit
is twice as large as that in the fiducial model.
These results demonstrate that merging of two massive GCs is possible
in massive dwarfs being destroyed by the Galaxy and  thus that 
anomalous GCs can originate from
such massive  dwarfs.

\section{Linking chemical evolution of dwarfs with
chemical abundances of GCs}

\subsection{One-zone models}

To validate the theoretical interpretation presented in the previous sections, 
we calculate the evolution of Ba (a $s$-process element),
Eu (a $r$-process one),  and Fe  abundances for dwarf galaxies with
different star formation histories.
The chemical evolution model which we adopt here is essentially the same as
those  constructed for the 
Fornax dSph
galaxy in T11. Although the star formation histories
of host dwarf galaxies for the Galactic GCs can be
quite diverse, we focus mostly on the host for M22.
We  slightly modify only two parameters 
(i.e., star formation rate, SFR, and  stellar initial mass
function) in the models by T11 
so as to be in better agreement with the chemical feature of the 
host dwarf galaxy for M 22. Since the details of the model is
given in T11,  we just briefly describe the essence of the models below.

The basis for the model is that a dwarf galaxy is formed through a continuous
  infall of 
almost pristine gas 
([Fe/H]$=-5$) from outside. With this framework, the star formation rate (SFR) 
at a time step $t$ is 
assumed to be proportional to the gas fraction with a constant rate 
coefficient $\nu$
Here, $\nu$ is the fraction of the gas mass that is converted 
into stars per Gyr. For the infall rate, we apply a formula that is 
proportional to $\exp(-t/\tau_{\rm in})$ with a 
timescale of 
infall of $\tau_{\rm in}$. For the best model, the 
values of ($\nu$, $\tau_{\rm in}$)=(1.0, 0.1) are assigned.
The initial mass function (IMF) is assumed to be a power-law mass spectrum with 
a slope of -1.35,  i.e.,  a canonical Salpeter IMF, 
with a fixed lower mass cutoff  of $0.05 M_\odot$ and
a variable upper mass cutoff of $m_{\rm upp}$.
The $r$-process yield from a 
neutron star merger is assumed as follows. 
We adopt the ejecta mass of a neutron star
merger is 0.01 $M_¥odot$. Since the frequency of neutron 
star mergers is estimated to be one per a few 1000 SNe II 
(Tsujimoto \& Shigeyama 2014),
we determine its frequency within the possible range 
so as to be the nucleosynthetic [Eu/Fe] ratio of +0.5 as observed in 
M22 or Galactic halo stars,
where the Fe yield is the average from an SN II:
In this case, the predicted (calculated) [Eu/Fe] exactly becomes +0.5.
A delay time of ejection from a neutron star merger is assumed to be $10^7$ yrs.
The details of  nucleosynthesis yields for Ba and light elements 
(C, N, and O) adopted in the present study
are  given in T11 and Tsujimoto \& Bekki (2011), respectively.

As in the  case for the Fornax, we find that a small $m_{\rm upp}$ of 25$M_\odot$ 
is the  best solution to obtain a high [Ba/Fe] ratio within a short 
timescale ($<1$ Gyr).
Therefore we mainly discuss the results of the best model with
($\nu$, $\tau_{\rm in}$, $m_{\rm upp}$)=(1.0, 0.1, 25) 
(referred to also as `high SFR and low $m_{\rm upp}$'
model) in the present study.
We do not consider the contribution from 
Type Ia SNe (SNe Ia) since no signature of chemical enrichment by SNe Ia 
is seen in the ratio of $\alpha$-elements to Fe for the two populations of M22. 
This assumption seems incompatible with the delay time distribution of SNe Ia obtained from the studies 
on the SN Ia rate in distant and nearby galaxies, which claim that SNe Ia start from a $\sim 10^8$ yrs delay from the initial star formation (e.g., Maoz et al. 2011). 
However, for instance, the Sagittarius dSph galaxy which is as massive as a candidate for the progenitor of M22, exhibits 
no signature of chemical enrichment by SNe Ia until [Fe/H]$\sim -1.3$ (de Boer et al. 2014a), where about 1-3 Gyr is implied to elapse after the 
start of star formation from the age-metallicity relation (McWilliam et al. 2013;
de Boer et al. 2014b).
In addition to the best case, we  show another 
two cases of ($\nu$, $\tau_{\rm in}$, $m_{\rm upp}$)=(0.3, 0.1, 25) 
(`low SFR and low $m_{\rm upp}$' model)
and (1.0, 0.1, 50) (`high SFR and high $m_{\rm upp}$' model)
to demonstrate the dependence of the results on the 
adopted parameters. Here we just see the early evolution of a 
dwarf galaxy for the duration of star formation of 1 Gyr, which is 
compatible with  the observed old ages of the Galactic GCs
(e.g., M22).

\subsection{Results}

Figure 8 shows the results for the best model corresponding to
the host dwarf of M22 and other two comparative ones.
The  best model denoted by the red solid line 
broadly reproduces the correlation of [Ba/Fe] and [Fe/H] 
corresponding to the two populations of M22 with a short 
timescale of a few $10^8$ yrs. Comparison of this model result with that for
the model with a high $m_{\rm upp}$ of 50$M_\odot$ (black dot-dashed line) 
reveals that an efficient Ba enrichment in comparison with Fe enrichment is 
realized by introducing a small $m_{\rm upp}$. 
It is attributed to the result that a small 
$m_{\rm upp}$ yields a reduction in the Fe mass ejected from each generation 
of SNe II, whereas the amounts of s-process elements synthesized in 
low-mass AGB stars remain unchanged.
It should be noted here that the time  evolution of [Ba/Fe] is rather rapid
(i.e., steeper age-dependence of [Ba/Fe])
for ages less than 0.5 Gyr and then it becomes less rapid 
(i.e., flatter [Ba/Fe]-age relation). 
This might have some constraints on the formation epochs of two GCs
that finally merge with other to form a single anomalous GC in dwarfs:
larger [Ba/Fe] differences in earlier formation of anomalous GCs via merging.

As shown in the upper panel of Figure 8,
the model with low SFR and low $m_{\rm upp}$ (blue short-dashed line)
has an almost same trend of [Ba/Fe] evolution  as obtained for the best model,
which suggests that  the rapidity of SF does not much influence
the evolution of [Ba/Fe]  in dwarf galaxies.
However, this model with low SFR  shows (i)  a rather low [Fe/H]
that is not consistent with [Fe/H] of  metal-poor GCs ([Fe/H]$\sim -1.6$)
and (ii) a very slow [Fe/H] evolution
that can not cause a significant [Fe/H] difference of $\sim 0.2$ dex
after 0.3 Gyr, as required for the case of the host of M22. 
These results of three models therefore clearly demonstrate that
both high SFR and lower $m_{\rm upp}$ are required for explaining both the increase
of [Ba/Fe] by $\sim 0.3$ dex and that of [Fe/H] by $\sim 0.15$ dex within
$\sim 0.3$ Gyr chemical evolution of dwarf galaxies.
It it our future work to investigate how such a rapid SF with lower $m_{\rm upp}$
is possible for GC-host dwarf galaxies.

The upper panel of Figure 9 shows the time evolution of [Eu/Fe] 
for the best model, which is assumed to be  "massive" 
dwarf galaxies such as the Fornax dwarf spheroidal galaxy
(see \S 5.3 for the discussion
on how the results depend on the stellar masses). 
In this model,  mergers between neutron stars
(`NS-NS merger') can occur  so that [Eu/Fe] can be 
kept constant (e.g., Tsujimoto \& Shigeyama 2014). However,
the evolution of [Eu/Fe] in low-mass dwarf galaxy models
can not be constant as this best model,
because NS-NS merger is an extremely rare event (i.e., almost no NS-NS mergers,
which end up with a decreasing trend of [Eu/Fe] with [Fe/H]).
The lower panel of Figure 9 shows that the time evolution of [Ba/Eu] 
is very similar to that of [Ba/Fe] in Figure 8, which is totally expected
for a flat evolution of [Eu/Fe] in massive dwarf galaxies 
(Tsujimoto \& Shigeyama 2014).
These results in Figure 9 suggest that  [Eu/Fe] and [Ba/Eu] of GCs
have fossil information of the chemical evolution of their host dwarfs.

Although the present one-zone chemical evolution model can reproduce the observed [Ba/Fe]
difference in the two populations of M22 reasonably well, it can not simply explain
the observed difference in [(C+N+O)/Fe] in the two populations (Fig. 17 in Marino et al. 2011).
Figure 10 shows that an enhancement of [C/Fe] by 0.1 dex is possible between ages of 0.2 and
0.8 Gyr in the  best model (indicated by the red solid line).
Figure 10, however,
shows that [(C+N+O)/Fe] can not become enhanced by +0.1 dex within 1 Gyr
in the three models,
which means that the present model fails to reproduce  both [Ba/Fe] and [(C+N+O)/Fe] spreads
observed in M22 self-consistently. Mixing of ISM with ejecta from AGB stars and SNe is 
assumed in the present model so that [(C+N+O)/Fe] can not be rapidly enhanced (within 300-500
Myr). Such a large difference in [(C+N+O)/Fe] between the two populations might be possible,
if new stars can form from gaseous ejecta from AGB stars 
without mixing 
so well with ISM.
We will investigate this issue in our future studies using new chemical evolution models
that can freely change the mass fraction of ISM mixing with AGB ejecta.

\section{Discussion}

\subsection{What distinguishes between
normal and anomalous GCs ?}

The present study has demonstrated that GCs with
internal metallicity spreads can be formed from 
merging of at least two GCs within their host dwarf
galaxies before the complete destruction of
the hosts by the Galactic tidal field.
In this merging scenario of anomalous GC formation,
less massive two GCs can not merge with each other quickly within dwarfs so that
they can not have internal metallicity spreads. 
Although such less massive  GCs can have FG and SG populations
with anti-correlations between light elements through self-enrichment
processes, they are much more likely to be stripped from
their hosts by the strong tidal field of the Galaxy.
Thus, the initial total mass of a GC  can be one of 
parameters that determine whether the GC can 
become an anomalous GC with  internal metallicity
spreads or normal GCs in this GC merging scenario.

Most of the Galactic anomalous GCs are observed
to show abundance spreads in $s$-process
elements and each of the  $s$-poor and $s$-rich sub-populations
have its own C-N and O-Na anti-correlations  
(s-Fe-anomalous GCs;  Marino et al. 2015).
In the GC merger scenario,  self-enrichment of intra-cluster gas
by AGB stars can proceed independently in each of two GCs formed
in different places  
with a time lag between two 
GC formation events ($t_{\rm lag}$).
Therefore, if merging of two GCs
can occur at least $\sim 300$ Myr after the younger of the two
was formed (i.e., $t_{\rm merge}> t_{\rm lag}+$300 Myr,
where 300 Myr corresponds roughly to the timescale of
self-enrichment),
then the new GC formed from GC merging
can naturally become  s-Fe-anomalous GCs.
It has long been suggested that self-enrichment of intra-cluster gas  by
Type II supernovae can cause internal metallicity variations in GCs,
if the GCs are massive enough to retain the SNII ejecta 
(e.g., Baumgardt et al. 2008).
In this in-situ self-enrichment scenario,  self-enrichment both by
SNII and AGB stars is possible in massive GCs, and accordingly,
$s$-Fe-anomalous GCs can be formed.
However it is not clear in this scenario how each of the $s$-rich
and $s$-poor populations in anomalous GCs (e.g., M22) 
show anti-correlations
between light elements abundances.

In the GC merging scenario,   
$t_{\rm lag}$ and the rapidity of star formation
in GC host dwarf ($t_{\rm sf}$) are the two main parameters
that control the degree of abundances spreads in [Fe/H] and
$s$-process elements of anomalous GCs.  If $t_{\rm lag}$ 
is shorter in a dwarf, then $\Delta$[Fe/H] and $\Delta$[Ba/Fe]
of anomalous GCs 
are smaller for a given dwarf's SFH in this scenario.
On the other hand, 
if a GC-host dwarf is forming stars slowly,
then $\Delta$[Fe/H] and $\Delta$[Ba/Fe] in anomalous GCs
formed in the dwarf  can be small
even for larger $t_{\rm lag}$. 
It should be noted here that 
$t_{\rm lag} < t_{\rm merge}$
and $t_{\rm sf} < t_{\rm merge}$ are required for
the formation of s-Fe-anomalous GCs.
It depends on the physical properties of GC-host dwarfs
whether these two conditions can be met.
The present results are based on purely collisionless simulations, which do not
incorporate possible important effects of baryon physics 
on the orbital evolution of GCs in low-mass dwarfs, such as strong supernova feedback
effects on  ISM in low-mass dwarf embedded in massive dark matter halos
(e.g., Maxwell et al. 2012). Since such bayonic effects 
can possibly influence $t_{\rm merge}$, $t_{\rm sf}$, and $t_{\rm lag}$,
we will need to investigate whether 
both $t_{\rm lag} < t_{\rm merge}$
and $t_{\rm sf} < t_{\rm merge}$ are satisfied in some  dwarf galaxies 
based on more sophisticated hydrodynamical simulations of dwarf galaxy formation.

\subsection{Where do metal-rich  anomalous stars come from ?}

Johnson et al. (2015) have recently made a detailed analysis of
chemical abundances of the Galactic GC NGC 6273 and found
that NGC 6273 has at least two stellar populations
with [Fe/H]
ranging from $-1.8$ to $-1.3$. In addition to the two populations,
the presence of a third metal-rich `anomalous stars' that
have chemical abundance patterns different from those
of the two populations have been discovered
in NGC 6273 (Johnson et al. 2015).
Using multi-wavelength HST photometry and previous 
results of chemical abundance studies for M2,
Milone et al.  (2015) have also found a third population (`C'
component; anomalous
stars) that does not have sub-population with different
abundances of light elements in M2.
It is well known that the Galactic GC $\omega$ Cen has such metal-rich stars
(e.g., Lee et al. 1999), which implies that the star formation history
of the GC should be quite different from those of other normal GCs.

It is unclear how these anomalous stars were formed 
after the formation of other two major populations in these massive GCs.
The present study has shown that if 
the mass-ratio of a GC to its host halo ($f_{\rm m, gc}$) is higher in
a massive dwarf galaxy, then field stars in the central region of the dwarf
can finally reside within the merger remnant of GCs after they
are tidally captured by the remnant. The present study therefore
suggest a new scenario in which
the observed metal-rich anomalous stars in NGC 6273, M2,
and $\omega$ Cen are
originally field stars in their GC host dwarf galaxies.
The higher [Fe/H] of anomalous stars
in these GCs can reflect the chemical enrichment
processes of their host GCs after the formation of the GCs in this 
scenario: it is not due to a prolonged SF {\it within} the GCs.
The gravitational capture of field stars by massive GCs are likely
to occur in the central regions of dwarfs,
where star formation can proceed very rapidly
owing to possible high-gas densities.
As shown in Figure 8 of this paper,
[Fe/H] can change very rapidly by an order of magnitude within a timescale of $\sim 0.3$ Gyr,
which is shorter than the timescale of GC merging.
It is therefore possible that nuclear regions have already high [Fe/H] when
GCs merge and consequently start to capture nuclear field stars.
Therefore, the third populations in these GCs
can be more metal-rich in this scenario.
The metal-rich third population in NGC 6273 exhibits a 
low level of $s$-process,
which suggests that in a central region of a dwarf galaxy,
a rapid SF likely proceeds with a short timescale, 
resulting in r-process-dominant abundances.

The new scenario predicts that  low-mass GCs, 
can not have metal-rich anomalous stars,
because such low-mass GCs 
are more likely to be tidally stripped from their host dwarfs
before they spiral-in the nuclear regions of their hosts.
Given that GCs can not become anomalous GCs with internal metallicity
spreads without GC merging,
the new scenario suggest that only anomalous GCs can have metal-rich
anomalous stars.
Although three GCs have been observed to have anomalous stars so far,
it is not observationally clear whether the presence or absence
of anomalous stars depend on the degree of internal metallicity 
variations in GCs.
The detailed theoretical predictions on the chemical abundances
and dynamical properties  (e.g., spatial distributions
within GCs)  of anomalous
stars in anomalous GCs are still lacking so that we can not discuss
whether the new scenario is really promising in explaining their
fundamental properties in a fully self-consistent manner.
More observational and theoretical studies are required for better
understanding the origin of these unique stellar populations
in GCs.

\subsection{GC hosts should be massive: A clue from Eu}

Marino et al. (2011) found that [Eu/Fe] in stars of the 
Galactic GC M22 do not depend on [Fe/H] for $-2.0<$[Fe/H]$<-1.6$. 
The observed flat evolution of [Eu/Fe] is consistent with GC host dwarfs 
being as massive as the Fornax dwarf galaxy or the LMC as follows. 
The constant [Eu/Fe] with an increasing [Fe/H] means that the Eu abundance 
increases with an increasing Fe abundance. Such a continuous increase of 
r-process element is characteristic of the chemical evolution of massive galaxies 
with their stellar masses larger than $\sim10^6 M_\odot$ such as the 
Fornax and the Sagitarrius dwarf galaxies (Tsujimoto \& Shigeyama 2014). 
In contrast, faint dwarf galaxies such as the Draco galaxy exhibit a 
constant [Eu/H] of $\sim -1.3$ with no apparent increase in the Eu abundance 
over the metallicity range of $-2<$[Fe/H]$<-1$. In other words, [Eu/Fe] 
decreases with an increasing [Fe/H] in faint dwarf galaxies. 

These results for low-mass dwarfs  imply that there are no 
Eu production events while more than $10^3$ SNe II increase the Fe abundance
in the ISM of the dwarfs.
This level of rarity of Eu production is compatible with the frequency expected 
for neutron star mergers, which is estimated to be one per 1000-2000 SNe II 
(Tsujimoto \& Shigeyama 2014). For instance, the Draco galaxy, the stellar mass of 
which is $\sim 3\times10^5 M_\odot$ (Martin et al.~2008) is likely to host  
$\sim 1500$ SNe II in total, assuming a canonical IMF. 
Thus it is no wonder that no neutron star merger has occurred in the 
Draco, which results in the observed decreasing [Eu/Fe] feature. 
On the other hand, the Fornax dwarf galaxy ($\sim 2\times 10^7 M_\odot$) 
is predicted to host $\sim 100$ events of neutron star mergers over the 
whole evolution. It will yield  a continuous Eu increase with time, which is equivalent to the constant [Eu/Fe] feature as observed.
Therefore, the constant [Eu/Fe] feature is the compelling 
evidence that  a host galaxy for M22 should be 
sufficiently massive to continuously host a rare event of r-process production.

\subsection{A possible unified scheme}

The present study suggests that there are at least
five key physical  parameters
that control the formation processes and chemical abundances of
GCs. The first parameter
is the total mass ($M_{\rm h}$) of a GC-host dwarf galaxy
which determines whether a GC can be formed from massive gas clouds in the
dwarf. Only if $M_{\rm h}$ of a dwarf galaxy
exceeds a threshold halo mass ($M_{\rm h, th}$),
FG stellar systems massive enough
to host SG stars can be formed in the dwarf, as recent numerical simulations
have demonstrated (Bekki 2016).
The second parameter
is the total mass of a GC ($m_{\rm gc}$), which controls
the self-enrichment process of intra-cluster gas in forming GCs.
If $m_{\rm gc}$ is larger than a threshold GC mass ($m_{\rm gc, th}$),
then gaseous ejecta from AGB stars or massive stars can be retained
and mixed with pristine gas
in the deep gravitational potential of the GC so that  new stars can form from
the chemically enriched gas. This threshold GC mass for the formation
of SG stars has already been discussed by several authors 
(e.g., D08; B11; Conroy \& Spergel 2011). 

The third parameter is the timescale of GC merging ($t_{\rm merge}$) 
within GC-host dwarfs.  If this $t_{\rm merge}$ is shorter than
the destruction 
timescale of GC host dwarf ($t_{\rm dest}$) by the Galactic tidal
field during dwarf accretion onto the Galaxy, then
anomalous GCs with metallicity spreads can form from GC merging. 
If $t_{\rm merge}$ is longer than $t_{\rm dest}$, then GCs are stripped
from their host dwarfs before GC merging
and consequently become the Galactic `normal' GCs.
Since the dynamical friction process of GCs within dwarfs is a key
determinant for $t_{\rm merge}$,   
$m_{\rm gc}$ and mass densities and kinematics of 
field stars can control $t_{\rm merge}$. 
More massive GCs are more likely to have
shorter dynamical friction timescale so that they can become
anomalous GCs.

The fourth parameter is $t_{\rm dest}$, which can control
the duration of star formation in dwarfs. Star formation can last longer
in more massive dwarfs, because they are more likely to be  disrupted  by
the Galactic tidal fields more slowly. 
If $t_{\rm dest}$ in  a massive dwarf
is as long as several Gyr, then chemical enrichment
in the central region can lead to a very large [Fe/H] spread
and a significant contribution of low-mass AGB stars to the evolution
of $s$-process elements. The stellar nucleus or nuclear cluster in
the dwarf can therefore show large abundance spread in different elements
(like $\omega$ Cen). The fifth parameter is $f_{\rm m, gc}$
(the mass-ratio of a GC to its host galaxy's dark matter halo),
which controls the capability of merged GCs to capture field stars. 
GCs in dwarfs with higher $f_{\rm m, gc}$ are more likely to
have field stars within them after GC merging,
and these field stars are not initially in GCs and later captured by
the GCs. These captured field stars may well be observed as anomalous
metal-rich stars in anomalous GCs.

Figure  11 illustrates how the five parameters determine the levels
of internal abundance spreads of GCs
and Table 4 summarizes the physical meaning of the
five parameters: this is just one of possible
unified scenarios of GC formation. There are only five different classes
of GCs in this illustration
and only one proto-type of each GC class is shown just for
clarity.    
In this scenario, the levels of internal abundance spreads in GCs 
depend not only on the physical properties of GCs ($m_{\rm gc}$)
but also on the chemical and dynamical histories of their host GCs.
This scenario is based on {\it separate} investigation of 
chemical and dynamical evolution of GCs and their host dwarfs,
and accordingly they are not fully self-consistent.
It is thus our future study to investigate how the chemical properties
of GCs depend on their formation processes and the evolution of their host 
dwarfs
using chemodynamical simulations of galaxy formation with a model for GC
formation.

\section{Conclusions}

Using dynamical simulations of dwarf galaxies with GCs and chemical
evolution models,
we have investigated (i) merging processes of GCs in their host dwarf galaxies,
(ii)  stripping of GCs from their host dwarfs,
(iii) tidal capture of field stars by merging GCs,
(iv) final kinematics of merged GCs,
and (v) various chemical abundances (e.g., [Fe/H] and [Ba/Fe]) of merged GCs.
The present dynamical simulation is better than BY12 in that 
both GCs and their host dwarf galaxy are represented by N-body particles.
Furthermore, the present study has considered that
the  chemical evolution of GC host dwarfs are linked to the internal
abundance spreads of GC stars.
The main results are summarized as follows. \\

(1) Merging between massive GCs with their masses ($M_{\rm gc}$) larger
than $3\times 10^5$ ${\rm M}_{\odot}$ is inevitable in their host dwarf galaxies
with $M_{\rm h}= [3 \times 10^9 - 3\times  10^{10}]$ ${\rm M}_{\odot}$
because of more efficient dynamical friction of the GCs against disk
field stars in the host dwarfs. Given that the time scale of GC merging
($t_{\rm merge}$) is less than a few Gyr in these massive dwarfs,
the merged GCs can become off-center nuclei before their hosts are 
completely destroyed by the Galactic tidal field.
These results do not depend on whether GCs have both two generations
(i.e.,  FG and SG)
of  stars or just single generation of stars. \\

(2) Low-mass GCs with $M_{\rm gc} \le 10^5$ ${\rm M}_{\odot}$
can not lose their orbital energy and angular momentum efficiently
so that they can not sink rapidly
into the central regions of their hosts.
Therefore  they
are more likely to be tidally stripped from
their host dwarfs before GC merging to become `normal' GCs
(with C-N and Na-O anti-correlations).
This result combined with the above result (1) strongly
suggests that GC masses can distinguish between normal GCs
and anomalous ones 
with [Fe/H] spreads. 
It has long been suggested that chemical enrichment of intra-cluster gas through
gas ejection from Type II supernovae might proceed more efficiently
in more massive GCs and consequently cause [Fe/H] spreads among GC stars.
The present study suggests that simple GC-GC merging can also equally
explain the observed [Fe/H] spreads in anomalous GCs. \\

(3) If the formation epochs
of  two GCs in dwarfs are separated by $\sim 300$ Myr, then
the chemical abundances of [Fe/H] and
[Ba/Fe] can be different
by  0.15 dex and 0.3 dex, respectively,  owing to chemical
enrichment by AGB stars and supernovae within the dwarfs.
However,  star formation in the GC-host dwarfs 
needs to proceed rather rapidly 
to obtain such rapid chemical enrichment,
and the upper-mass cutoff ($m_{\rm upp}$) of IMF
can influence the chemical evolution.
The internal abundance spreads of merged GCs depend both on the 
star formation histories of their host dwarfs and on the time lag
between two GC formation events.
We suggest that star formation histories of dwarfs hosting anomalous
GCs can be different from dwarfs hosting  normal GCs.
The observed no/little dependence of [Eu/Fe] on [Fe/H] in M22 is consistent
with a scenario in which M22 was formed in a massive dwarf galaxy
at a  high redshift. \\

(4) Field stars can be captured by merged GCs in the central regions
of GC-host dwarfs, if the mass-ratios of GCs to their hosts are large
($M_{\rm gc}/M_{\rm h} \sim 0.0001$). The chemical abundances of these
field stars should be different from those of stars initially in
merging GCs. We suggest that metal-rich `anomalous stars' observed
in some GCs (e.g., NGC 6273; Johnson et al. 2015) can be formed
as a result of tidal capture of field stars by merging GCs within
dwarfs. As already discussed in BY12,  stellar halos can be formed
around  merged GCs after tidal destruction of their host dwarfs. \\

(5) GC merger remnants inevitably show global rotation,
because orbital angular momentum of merging GCs is converted into
internal rotation of the remnants. 
The present study has investigated only merging of GCs with initially
no net global rotation (i.e., dense stellar systems dynamically
supported only by velocity dispersion) and found that the rotation
amplitude is only $\sim 1.5-6$ 
km s$^{-1}$ for $M_{\rm gc}=10^6$ ${\rm M}_{\odot}$.
The stellar kinematics of GC merger remnants could depend on
the initial stellar kinematics of merger progenitor GCs, which should be
investigated in our future works. \\

(6) The details of chemical abundances in anomalous GCs can depend on
the time lag between two GC formation events,
the timescale of GC host destruction,  and the chemical evolution processes
of GC host dwarfs,
if they were formed mainly from GC merging.
The time lag between the formation epochs of two GCs that finally merge to form
a single GC is very short for the case of NGC 1851 with
little/no [Fe/H] spread whereas it should be
at least a few Myr in the merger scenario of anomalous GCs with 
significant [Fe/H] spreads: GC merging alone does not necessarily
form single anomalous GCs in this scenario.
$\omega$ Cen could possibly have experienced both multiple GC merging and 
gas inflow from the outer part of its host dwarf for a longer time
scale so that it could have a very large [Fe/H] spread.
The unique history of $\omega$ Cen needs to be discussed in detail by
a separate paper.\\

\acknowledgments
We are   grateful to the referee for his/her constructive and
useful comments.
T. T. is assisted in part by JSPS KAKENHI Grant Number 15K05033.

\begin{deluxetable}{llllllllll}
\footnotesize  
\tablecaption{ Description of the parameter values
for the representative four models.
\label{tbl-1}}
\tablewidth{-2pt}
\tablehead{
\colhead{  ID  \tablenotemark{a}} &
\colhead{  $m_{\rm FG,1}$  \tablenotemark{b}} &
\colhead{  $a_{\rm FG,1}$ \tablenotemark{c}} &
\colhead{  $m_2$  \tablenotemark{d}} &
\colhead{  $R_{\rm gc, 1}$  \tablenotemark{e}} &
\colhead{  $R_{\rm gc, 2}$  \tablenotemark{f}} &
\colhead{  Dwarf  \tablenotemark{g}} &
\colhead{  Orbit  \tablenotemark{h}} &
\colhead{  $t_{\rm merge}$   \tablenotemark{i}} &
\colhead{  Comment \tablenotemark{j}} }
\startdata
M1 & 1.0 & 10.0 & 1.0 & 100 & 200 & D1 & O1 & 0.31 & Fiducial \\
M2 & 1.0 & 10.0 & 1.0 & 200 & 400 & D1 & O1 & 0.30 & \\
M3 & 1.0 & 10.0 & 1.0 & 400 & 600 & D1 & O1 & 1.33 &\\
M4 & 1.0 & 10.0 & 1.0 & 600 & 800 & D1 & O1 & N &\\
M5 & 0.3 & 5.8 & 1.0 & 100 & 200 & D1 & O1 & N & \\
M6 & 0.1 & 3.2 & 1.0 & 100 & 200 & D1 & O1 & N & \\
M7 & 1.0 & 10.0 & 0.1 & 100 & 200 & D1 & O1 & N & Unequal-mass \\
M8 & 1.0 & 1.0 & 1.0 & 100 & 200 & D2 & O1 & 0.52 &  FG-only\\
M9 & 0.3 & 5.8 & 1.0 & 100 & 200 & D2 & O1 & 1.34 & \\
M10 & 1.0 & 10.0 & 1.0 & 100 & 200 & D2 & O1 & 0.18 &  \\
M11 & 1.0 & 10.0 & 1.0 & 100 & 200 & D3 & O1 & N & \\
M12 & 1.0 & 10.0 & 1.0 & 100 & 200 & D1 & O1 & 0.41 & FG-only  \\
M13 & -- & --  & -- & -- & -- & D4 & O1 &  -- & No GCs \\
M14 & 1.0 & 10.0 & 1.0 & 100 & 200 & D1 & O3 & 1.73 &Young MW \\
M15 & 0.3 & 5.8 & 1.0 & 100 & 200 & D1 & O3 & 0.74 & \\
M16 & 1.0 & 10.0 & 1.0 & 100 & 200 & D1 & O4 & 0.54 & \\
M17 & 1.0 & 10.0 & 1.0 & 100 & 200 & D5 & O1 & 0.32 & SB profile \\
M18 & 1.0 & 10.0 & 1.0 & 100 & 200 & D1 & O2 & 1.25 & Distant orbit \\
M19 & 1.0 & 10.0 & 1.0 & 100 & 200 & D6 & O1 & -- & No GC \\
\enddata
\tablenotetext{a}{
The host dwarf galaxy model (e.g., D1) and orbital model (e.g. O1)  
are given in Table 2 and 3, respectively. The fiducial model adopts
D1, and O1, and NFW dark matter model.
The mass-ratio of SG to FG stellar systems are 0.2 for all models
with FG+SG systems.  GCs with only FG systems are clearly indicated
as `FG-only' in the comment column. 
}
\tablenotetext{b}{
The initial total mass of  FG stars of GC1 in units of $10^6 M_{\odot}$.
}
\tablenotetext{c}{
The scale-length of a FG systems ($a_{\rm FG,1}$) in a GC1.
The scale-length of the SG is $0.2 a_{\rm FG,1}$.
}
\tablenotetext{d}{
The mass-ratio of GC2 to GC1.
}
\tablenotetext{e}{
The initial distance of GC1 from the Galactic center in units of pc.
}
\tablenotetext{f}{
The initial distance of GC2 from the Galactic center in units of pc.
}
\tablenotetext{g}{
The host dwarf galaxy model.
}
\tablenotetext{h}{
The orbital evolution model around the Galaxy.
}
\tablenotetext{i}{
`N' means that two GC1 and GC2 can not merge
with each other before the completion of tidal destruction of
their host dwarf galaxy. The value given in each model is
$t_{\rm merge}$ in units of Gyr (the timescale of GC merging). '$-$' means 
that GC merging can not be defined owing to no GCs in the model.
}
\tablenotetext{j}{
Comments on each model, if any.
}
\end{deluxetable}

\begin{deluxetable}{lllllll}
\footnotesize  
\tablecaption{ Description of the parameter values
for GC-host dwarf galaxies.
\label{tbl-1}}
\tablewidth{-2pt}
\tablehead{
\colhead{  Model ID  \tablenotemark{a} } &
\colhead{  $M_{\rm h}$ \tablenotemark{b} } &
\colhead{  $r_{\rm vir}$ (kpc) \tablenotemark{c}} &
\colhead{  $M_{\rm s}$  \tablenotemark{d}} &
\colhead{  $r_{\rm s}$ (kpc) \tablenotemark{e}}  &
\colhead{  DM profile \tablenotemark{f}}  &
\colhead{  GC \tablenotemark{g} } }
\startdata
D1 & 1.0 & 9.2 & 1.5 & 1.3 & NFW & Yes \\
D2 & 0.3 & 5.0 & 0.45 & 0.7 & NFW & Yes \\
D3 & 3.0 & 15.9 & 4.5 & 2.3  & NFW & Yes \\
D4 & 1.0 & 9.2 & 1.5 & 1.3 & NFW & No \\
D5 & 1.0 & 9.2 & 1.5 & 1.3 & SB & Yes \\
D6 & 0.01 & 0.92 & 0.015 & 0.13 & NFW & No \\
\enddata
\tablenotetext{a}{
These model IDs are used in Table 1.
}
\tablenotetext{b}{
The initial total  mass of the dark matter halo  ($M_{\rm h}$)
of a GC-host dwarf
in units of $10^{10} M_{\odot}$.
}
\tablenotetext{c}{
The virial radius of the dark matter halo of a GC-host dwarf.
in units of kpc.
}
\tablenotetext{d}{
The initial total  mass of the stellar disk of a GC-host dwarf
in units of $10^8 M_{\odot}$.
}
\tablenotetext{e}{
The initial size of the stellar disk of a GC-host dwarf
in units of kpc.
}
\tablenotetext{f}{
NFW and SB represents the NFW and SB profiles, respectively, for the
initial radial density distributions of dark matter halos.
}
\tablenotetext{g}{
`Yes' (`No') means that a dwarf galaxy (does not) have GCs.
}
\end{deluxetable}

\begin{deluxetable}{llllll}
\footnotesize  
\tablecaption{ Model parameters for 
the three-component Galactic potentials and dwarf orbits.
\label{tbl-1}}
\tablewidth{-2pt}
\tablehead{
\colhead{  Model ID  \tablenotemark{a} } &
\colhead{  $M_{\rm disk}$ \tablenotemark{b} } &
\colhead{  $M_{\rm bulge}$ \tablenotemark{c}} &
\colhead{  $v_{\rm halo}$ \tablenotemark{d}} &
\colhead{  $R_{\rm i}$ \tablenotemark{e}} &
\colhead{  $f_{\rm v}$  \tablenotemark{f}} }
\startdata
O1 & 10.0 & 3.4 & 131.5 & 17.5 & 0.5 \\
O2 & 10.0 & 3.4 & 131.5 & 35.0 & 0.5 \\
O3 & 1.0 & 0.34 & 93.0 & 17.5 & 0.5 \\
O4 & 1.0 & 0.34 & 93.0  & 8.8 & 0.7 \\
\enddata
\tablenotetext{a}{
The first two models (O1 and O2) correspond to the present Galaxy
whereas other two (O3 and O4) mimic the early formation phase
of the Galaxy (before/during the formation of the first thin disk).
}
\tablenotetext{b}{
The initial stellar disk mass of the Galaxy in  units of
$10^{10} M_{\odot}$. 
}
\tablenotetext{c}{
The initial stellar bulge mass of the Galaxy in  units of
$10^{10} M_{\odot}$. 
}
\tablenotetext{d}{
The velocity parameter 
in the adopted logarithmic potential of the Galactic halo
in units of km s$^{-1}$.
}
\tablenotetext{e}{
The initial distance of a dwarf from the Galactic center in units
of kpc.
}
\tablenotetext{f}{
The circular velocity factor: $f_{\rm v}=1$ means that the velocity
of a dwarf is the same as the circular velocity at its initial location.
}
\end{deluxetable}

\newpage

\begin{deluxetable}{ll}
\footnotesize  
\tablecaption{Five key parameters that determine 
the levels of chemical abundance spreads in GCs.
\label{tbl-1}}
\tablewidth{-2pt}
\tablehead{
\colhead{  Key parameter  \tablenotemark{a} } &
\colhead{  Description  \tablenotemark{b} } }
\startdata
$M_{\rm h, th}$  &  A threshold host halo mass for GC formation\\
$m_{\rm g, th}$  &  A threshold GC mass for GCs with multiple stellar
populations\\
$t_{\rm merge}$  &  Timescale of GC merging\\
$t_{\rm dest}$  &  Timescale of dwarf destruction by the Galactic
tidal field\\
$f_{\rm m, gc}$  &  The mass-ratio of one GC to its host halo
\enddata
\tablenotetext{a}{
These are {\it possible} five key parameters that are used to
divide GCs into different types (i.e., different degrees of
abundance inhomogeneity). There would be other important
parameters for the origin of multiple stellar populations in GCs. 
}
\tablenotetext{b}{
The details of the physical meanings of these parameters
are given and discussed in the main text.
}
\end{deluxetable}

\newpage

\begin{figure}
\epsscale{0.9}
\plotone{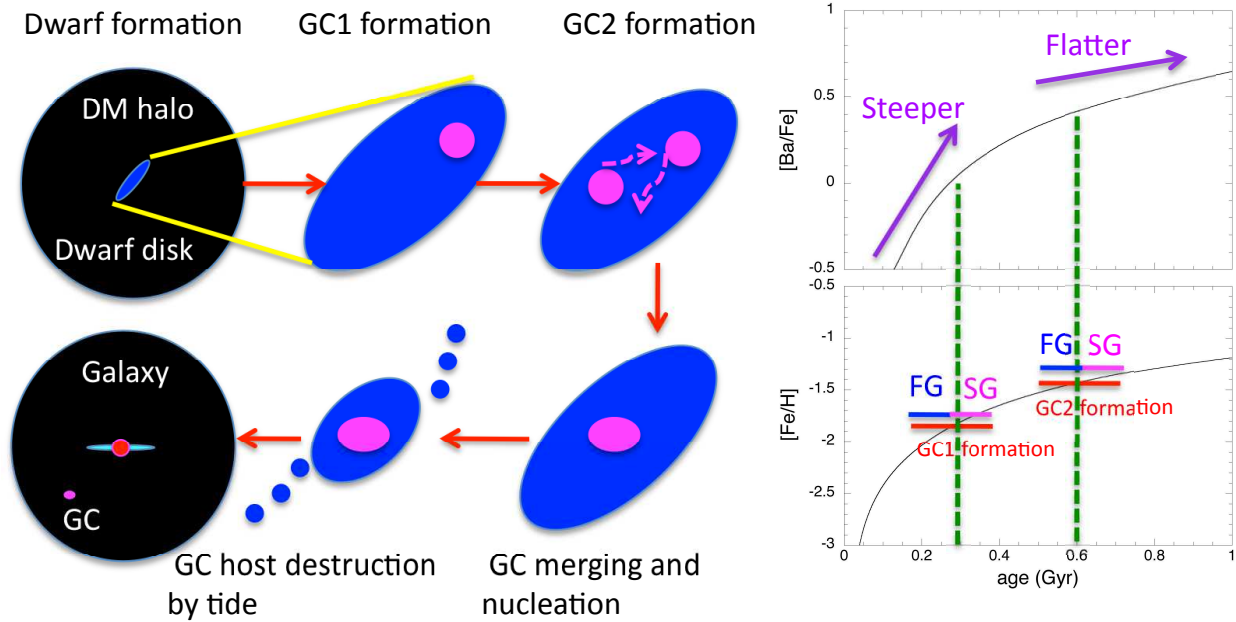}
\figcaption{
An illustrative figure for the `GC merging' scenario for the
origin of anomalous GCs with metallicity spreads. In this scenario,
two GCs (GC1 and GC2) were formed from gas clouds within 
a dwarf galaxy embedded in a massive dark matter halo at different
epochs. The two GCs lose orbital energy and angular momentum
due to dynamical friction against field star of the dwarf 
so that they can spiral into the nuclear region of the dwarf.
The two GCs consequently merge with each other to form an off-center
nucleus (or new nuclear GC). 
The dwarf galaxy can be completely destroyed by the tidal field
of the Galaxy during its accretion onto the Galactic halo.
The new GC can survive from the tidal destruction, and thus is identified
as a Galactic halo GC.
Each of the two GCs have FG and SG stars, the two GCs have different
[Ba/Fe] and [Fe/H] because they are formed at different epochs.
If the formation epochs of the two GCs are separated by $\sim 300$
Myr, then [Fe/H] and [Ba/Fe] can be different by 0.2 dex and 0.3
dex, respectively, between the two GCs as a result of chemical
evolution of the dwarf.  Therefore,  the new GC formed
from merging of the two GCs should have abundance spreads both in
[Fe/H] and [Ba/Fe]. The [Ba/Fe]-[Fe/H] relation becomes flatter
at later times, which can be reflected on the abundance patterns
of massive GCs such as $\omega$ Centauri.
\label{fig-1}}
\end{figure}

\newpage

\begin{figure}
\epsscale{1.0}
\plotone{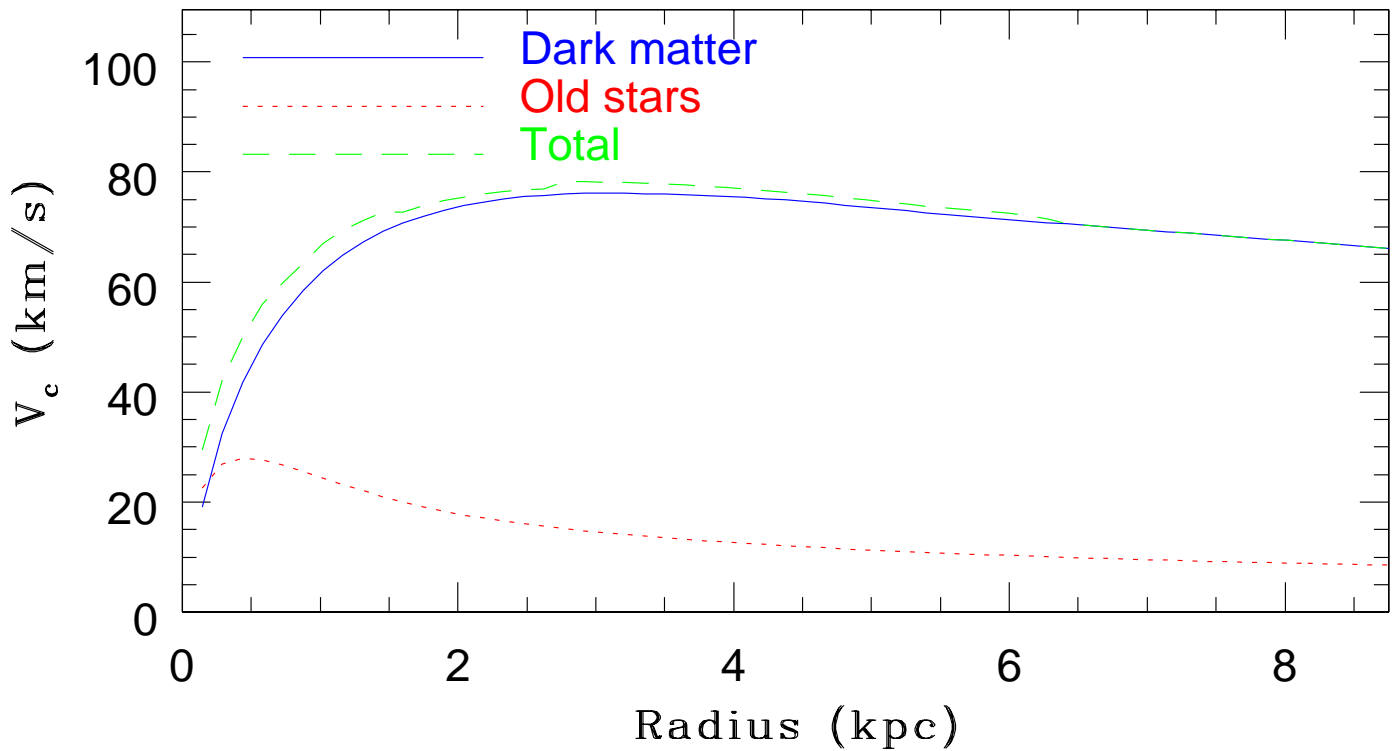}
\figcaption{
The contributions of dark matter (blue solid),  old stars (red dotted),
and all components (green dashed) to the circular velocity curve of
the dwarf disk galaxy model D1.
\label{fig-2}}
\end{figure}

\newpage
\begin{figure}
\epsscale{1.0}
\plotone{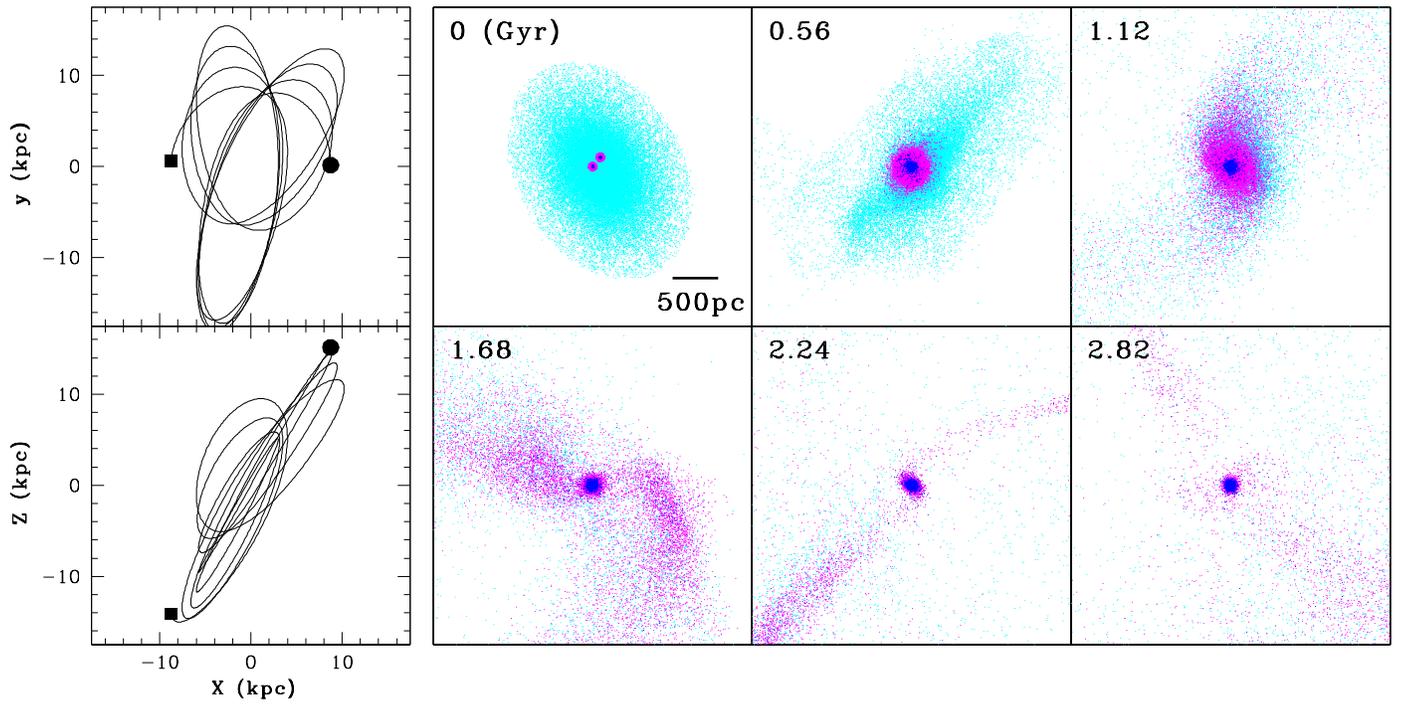}
\figcaption{
The left two panels show the orbit of  a dwarf projected
onto the $x$-$y$ (upper) and $x$-$z$ planes (lower) for the orbital
model O1. The initial and final locations of the dwarf are indicated
by filled circles and squares, respectively.
 The right six panels show the time evolution of the mass
distributions of the dwarf and its two GCs projected
onto the $x$-$y$ in the fiducial model
M1 with O1 orbital model. Disk (old) stars,  FG stars, and SG stars
are shown by cyan, magenta, and blue, respectively. The time $T$ that
has elapsed since simulation started is shown in the upper left corner
of each panel.  Only one in ten disk particles is shown so that
the file size of this figure  can be 
significantly reduced  (yet the key results can be 
clearly seen).
\label{fig-3}}
\end{figure}

\begin{figure}
\epsscale{1.0}
\plotone{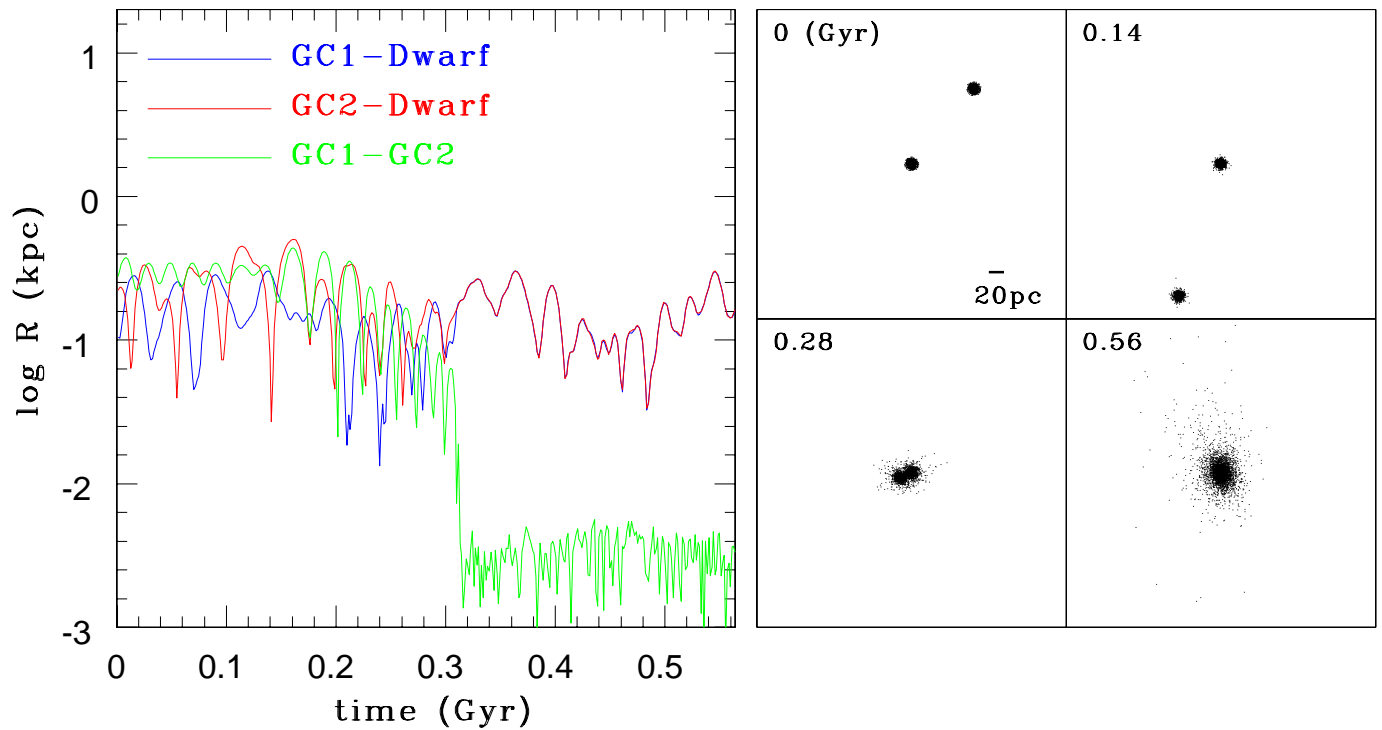}
\figcaption{
The bigger left panel shows the time evolution
of the distance ($R$) between
a GC-host dwarf and GC1 (blue), the host and GC2 (red),
and GC1 and GC2 (green) in the fiducial model M1. The smaller
four right panels describe the mass distribution of SG stars
in GC1 and GC2 projected onto the $x$-$y$ plane in the fiducial
model at our selected time steps. The epoch of GC-merging 
corresponds to the point
where the GC1-GC2 distance becomes rather small ($R<3$ pc)
suddenly and dramatically (i.e., $T \sim 0.3$ Gyr).
The `center' of the dwarf is defined as the position of the nuclear
particle, and the GC-dwarf distance is measured using the position
of each GC and that of the nuclear particle. Since the nuclear particle
is moving in the central region of the dwarf, the GC1-dwarf distance
can be still large even after the merged GCs can spiral in the nuclear
region of the dwarf.
\label{fig-4}}
\end{figure}

\begin{figure}
\epsscale{0.6}
\plotone{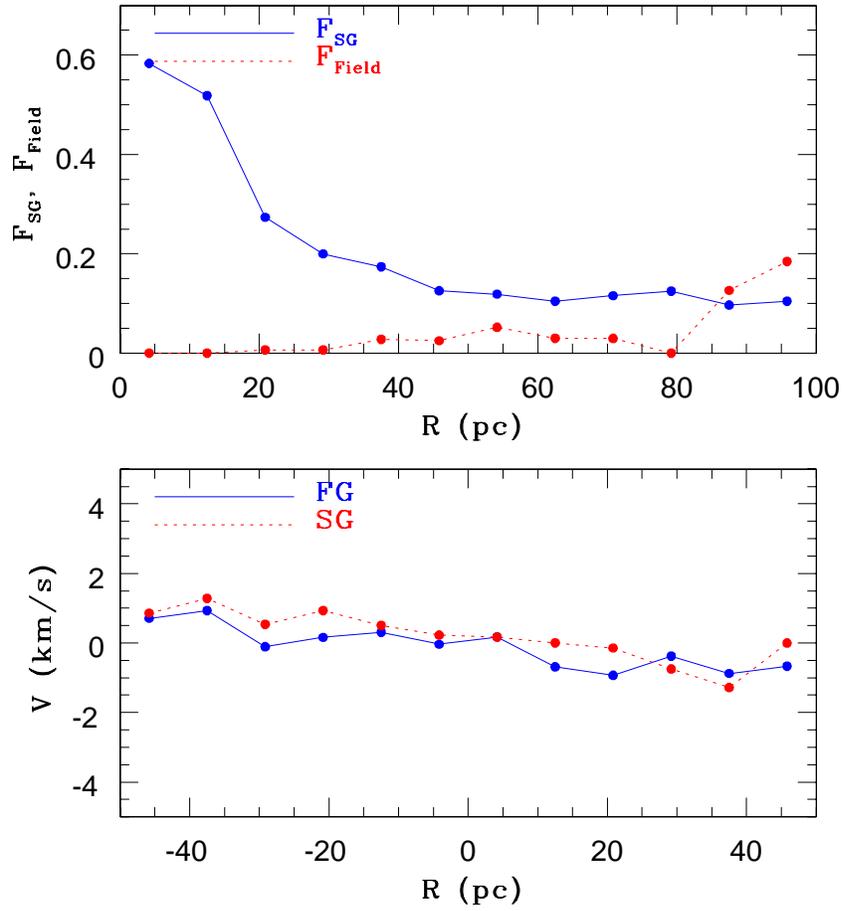}
\figcaption{
The upper panel shows
the radial dependence of the mass-ratio of SG to FG stars ($F_{\rm SG}$;
blue solid) and that of field to FG stars ($F_{\rm FG}$) in the 
new GC formed from major GC merging in the fiducial model M1
The lower panel shows the rotation curve profiles along
the $y$-axis for FG (blue solid)
and SG stars (red dashed) in the new GC. 
\label{fig-5}}
\end{figure}

\begin{figure}
\epsscale{1.0}
\plotone{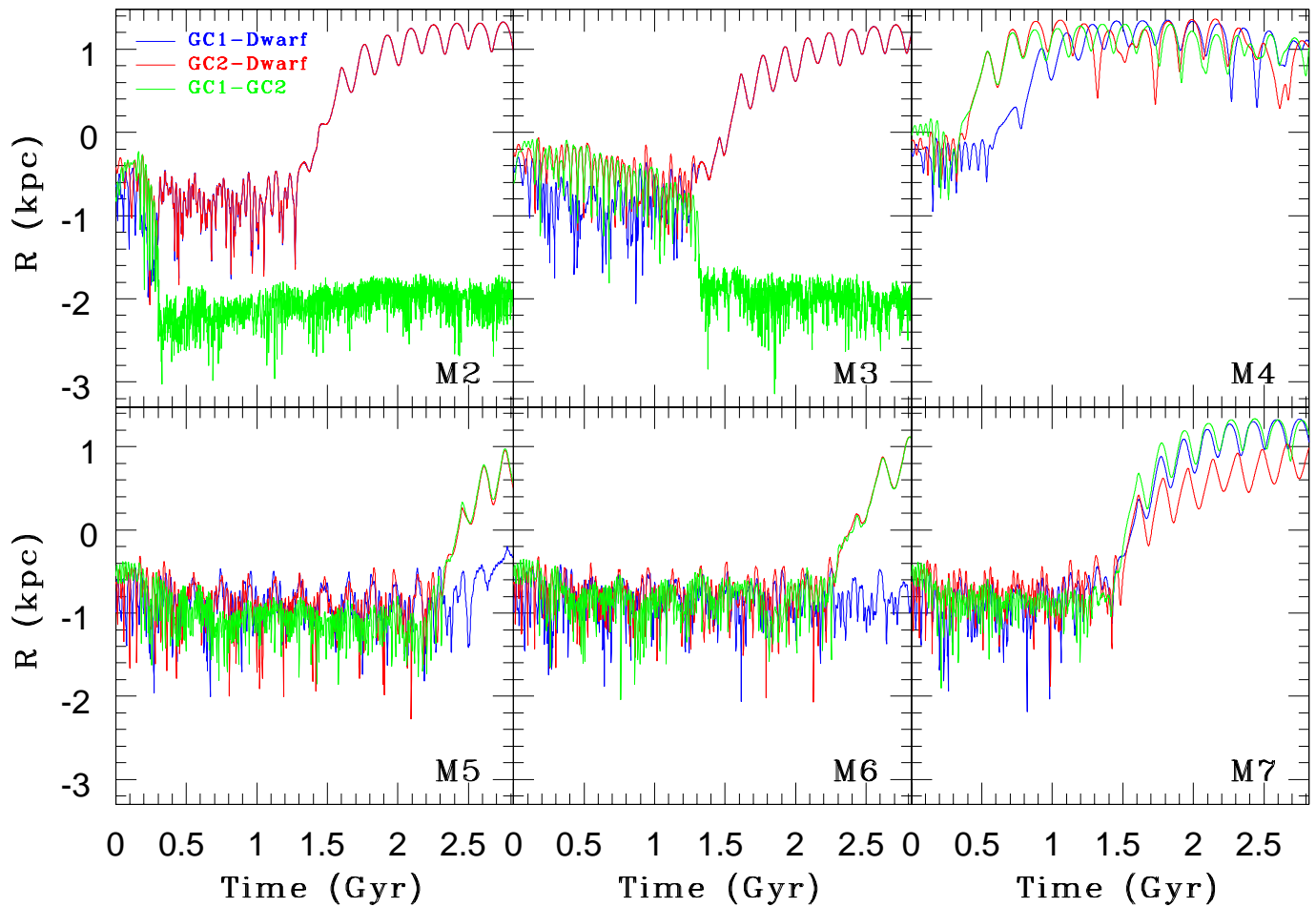}
\figcaption{
The same as the left panel of Figure 4 but for 6 different models
(M2-M7). The epoch when the merged GC (or either/both of the two GCs)
is stripped from the host dwarf galaxy corresponds to the point
where the GC-dwarf distance (red or blue line) become suddenly
large ($R>1$ kpc). Clearly, GC merging is possible in M2, M3,
and M5, whereas it is not in M6 and M7, in which at least one 
of the two GCs has a low mass ($m_{\rm FG, 1}=10^5 {\rm M}_{\odot}$). 
\label{fig-6}}
\end{figure}

\begin{figure}
\epsscale{1.0}
\plotone{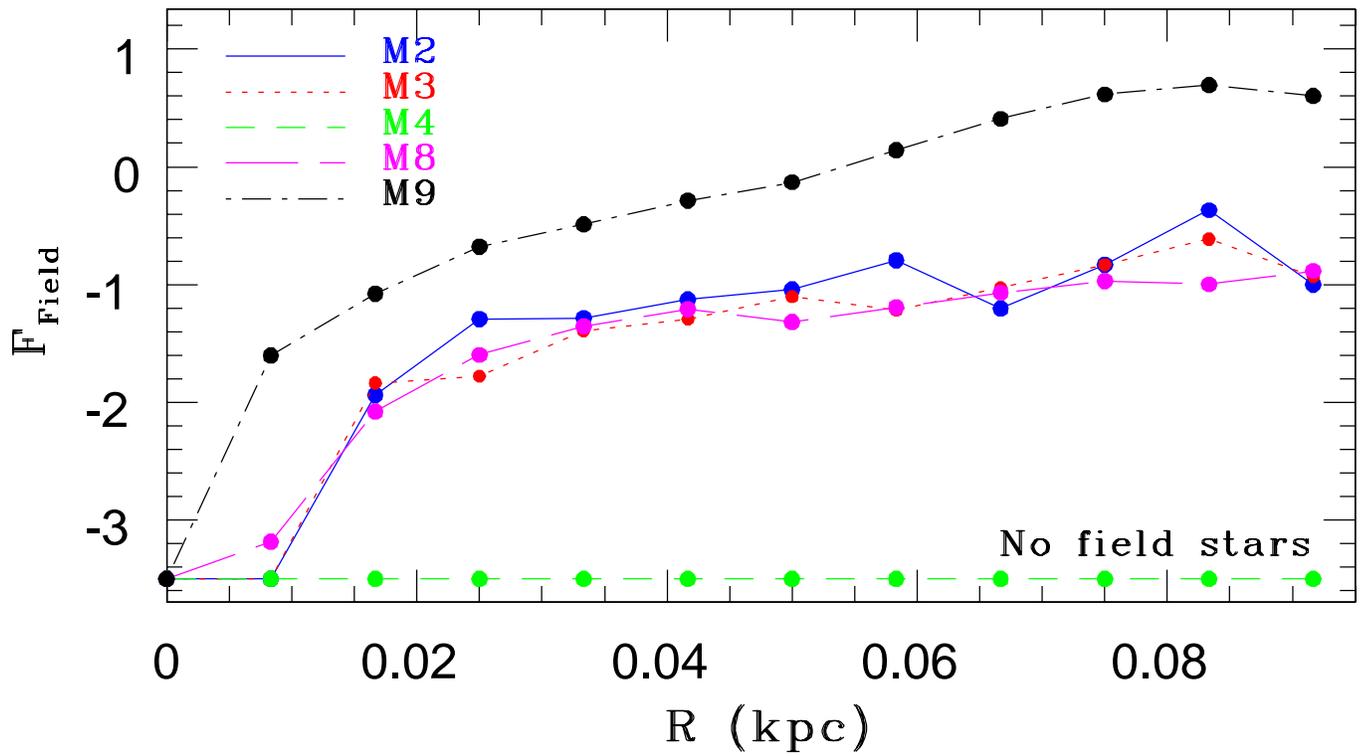}
\figcaption{
The mass-ratio of field to FG stars ($F_{\rm Field}$) for different models,
M2 (blue solid), M3 (red dotted), M4 (green dashed), M8 (magenta
long-dashed), and M9 (black dot-dashed).
For convinience,  $\log F_{\rm Field}=-3.5$ means no field stars around
the GCs. Accordingly, the model M4 does not show any field stars
in the halo of the GC.
\label{fig-7}}
\end{figure}

\begin{figure}
\epsscale{0.6}
\plotone{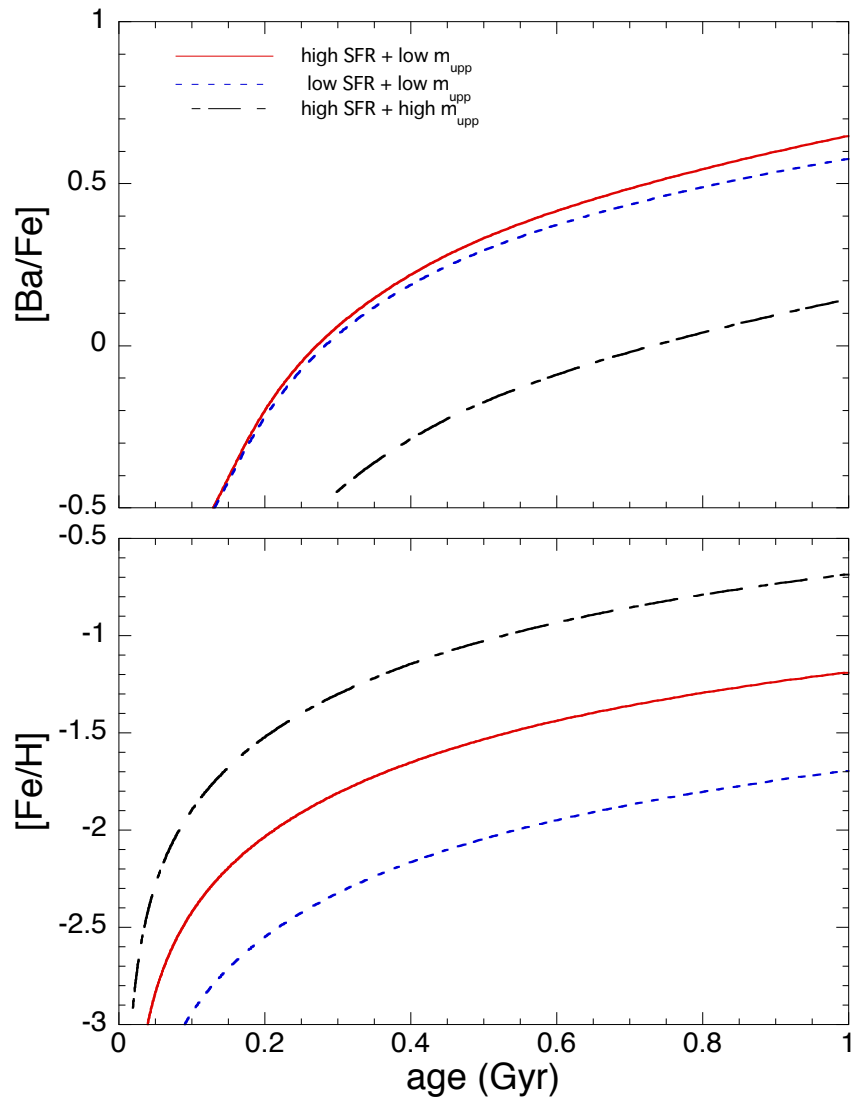}
\figcaption{
The time evolution of [Ba/Fe] (upper) and [Fe/H] (lower)
in the three chemical evolution models with different star formation
histories and upper cut-off stellar masses of the adopted IMF ($m_{\rm upp}$):
high star formation rate with low $m_{\rm upp}$ (red solid),
low star formation rate with low $m_{\rm upp}$ (blue short-dashed),
and high star formation rate with low $m_{\rm upp}$ (black dot-dashed).
The physical meanings of 'low' and 'high' star formation rate models 
are given in the main text.
\label{fig-8}}
\end{figure}

\begin{figure}
\epsscale{0.6}
\plotone{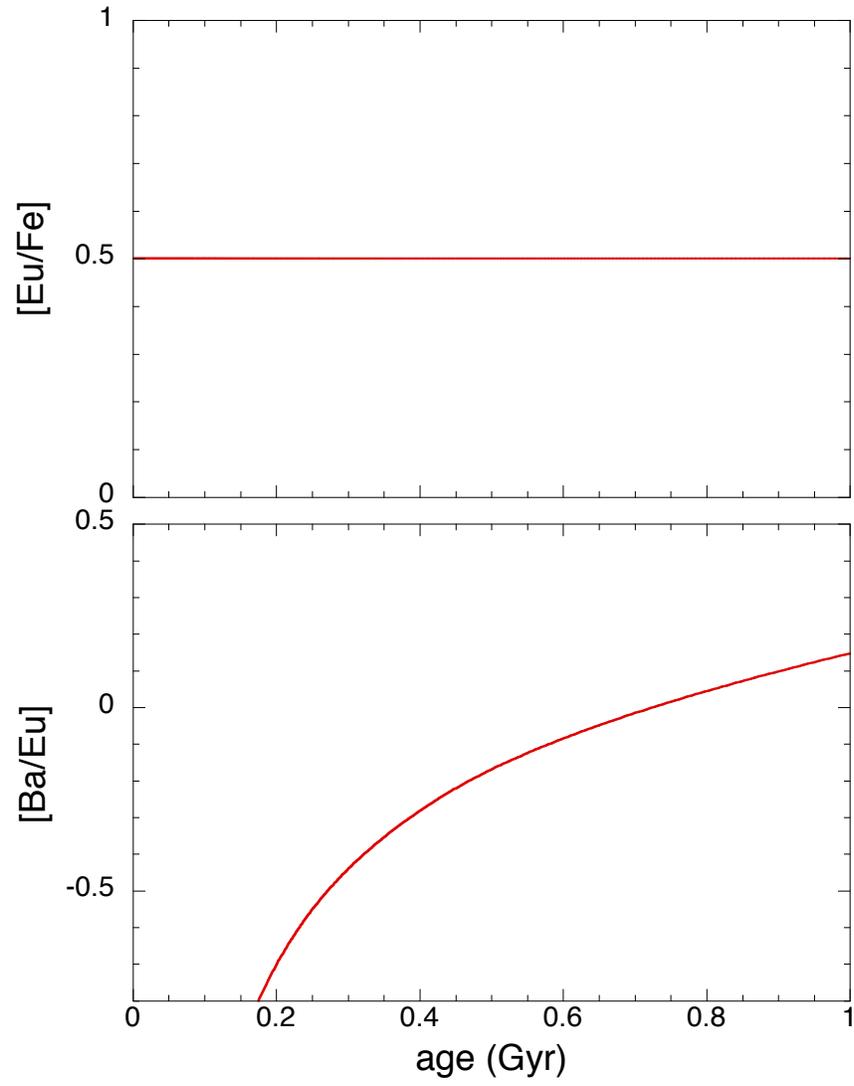}
\figcaption{
The same as Figure 8 but for [Eu/Fe] (upper) and [Ba/Eu] (lower) for
the best model. 
The implication of the flat [Eu/Fe] evolution
is given in the main text.
\label{fig-9}}
\end{figure}

\begin{figure}
\epsscale{0.6}
\plotone{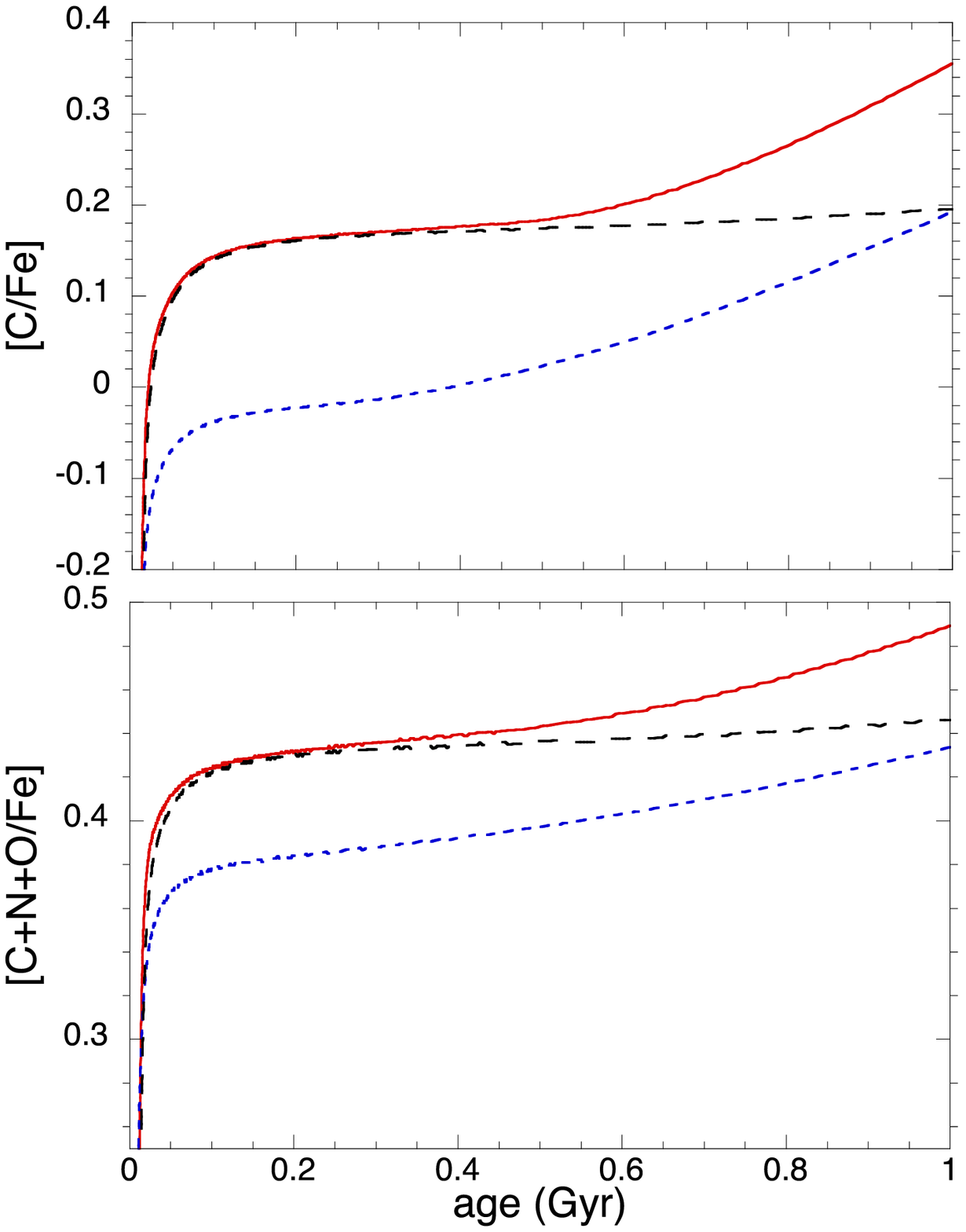}
\figcaption{
The same as Figure 8 but for the evolution of C (upper)  C+N+O (lower) for
the three models. 
\label{fig-10}}
\end{figure}

\begin{figure}
\epsscale{0.9}
\plotone{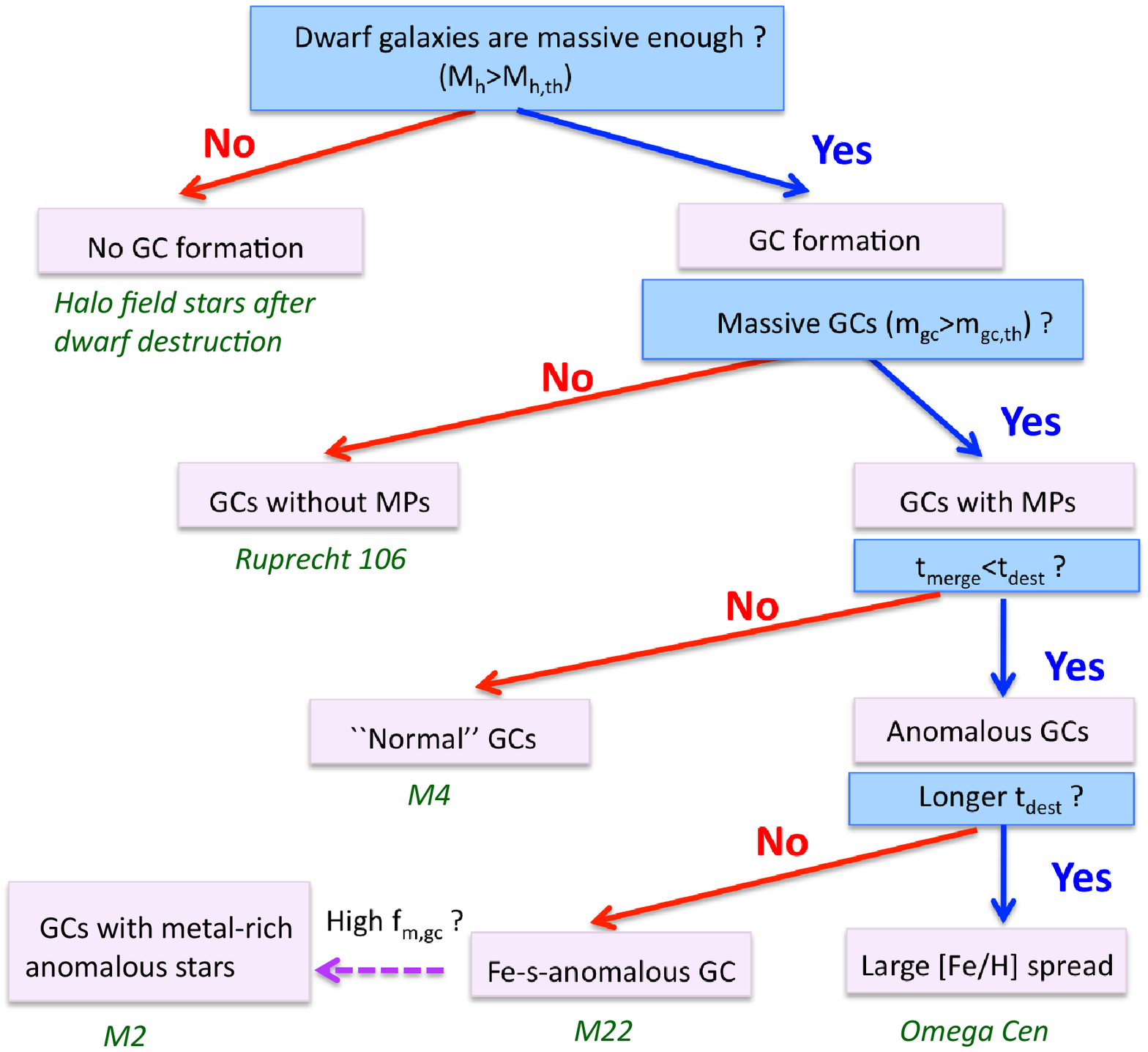}
\figcaption{
A schematic diagram for an unified picture of GC formation
with multiple stellar populations (MPs). There are five
key parameters in this diagram, $M_{\rm h, th}$ (threshold
dwarf mass for GC formation), $m_{\rm gc, th}$) (threshold
GC mass for GC formation with MPs), $t_{\rm merge}$ (timescale
for GC merging), $t_{\rm dest}$ (timescale of dwarf destruction),
and $f_{\rm m, gc}$ (mass ratio of a GC to its host halo).
GCs can be divided into different categories (e.g., GCs without MPs
and anomalous GCs) according to whether or not they meet the four specific
criteria (e.g., `$t_{\rm merge} < t_{\rm dest}$').
It should be stressed here that merging of two GCs does not necessarily
lead to the formation of a single anomalous GCs, 
because the two GCs need to form
at different epochs within their host dwarf.
Both M22 and NGC 1851 can be formed from GC merging, but the difference
in the formation epochs between two GCs for NGC 1851 is small. Accordingly,
NGC 1851 can not become  an anomalous GC with a significant [Fe/H]
spread.
The extensive discussion on GC formation
based on this diagram is given in the main text.
\label{fig-10}}
\end{figure}

\appendix
\section{Possible influences of stellar galactic nuclei on the timescale of 
GC merging}

Although stellar galactic nuclei (SGN) are included in the initial dwarf disk models by BY12,
they are not modeled in the present study. The initial presence of SGN might be able
to influence the GC merging processes in dwarfs significantly. We have therefore investigated
how SGN can influence the orbits of two GCs in the fiducial model by assuming that
the initial masses of SGN ($M_{\rm nuc}$) are either $10^5 M_{\odot}$ or $10^6 M_{\odot}$. 
The SGN is modeled by a point-mass particle with $M_{\rm nuc}$ rather than by
a self-gravitating system in this investigation.
Given that the initial stellar mass in the fiducial model ($M_{\rm s}$)
is $1.5 \times 10^8 M_{\odot}$,
the adopted $M_{\rm nuc}$ ($0.15-1.5$\% of $M_{\rm s}$) is quite reasonable. Figure 12
shows that the  GC merging timescale ($t_{\rm merge}$) is not so different between
these two models with SGN and the fiducial one (M1). The two GCs in these models merge
with each other within $0.7$ Gyr ($t_{\rm merge}=0.45$ Gyr for $M_{\rm nuc}=10^6 M_{\odot}$
and $t_{\rm merge}=0.64$ Gyr for $M_{\rm nuc}=10^5 M_{\odot}$).
The presence of SGN can slightly lengthen GC merging (by $0.1-0.3$ Gyr depending $M_{\rm nuc}$)
owing to dynamical interaction between SGN and each of the two GCs.
We therefore suggest that $t_{\rm merge}$  can not be significantly influenced by
the presence of SGN in dwarf disk galaxies.

\begin{figure}
\epsscale{0.8}
\plotone{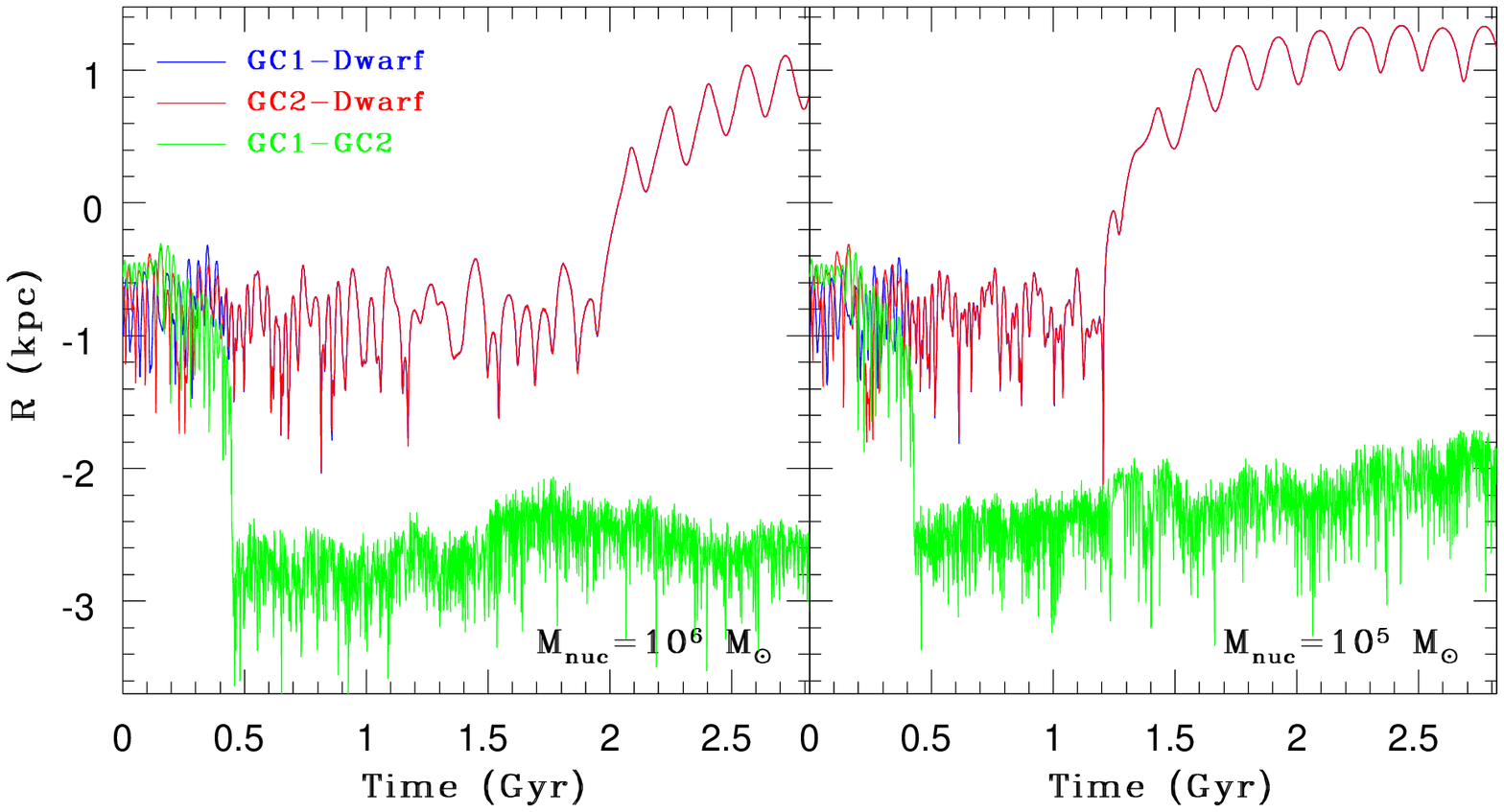}
\figcaption{
The same as Figure 6 but for the two models with $M_{\rm nuc}=10^6 M_{\odot}$ (left)
and $M_{\rm nuc}=10^5 M_{\odot}$ (right), $M_{\rm nuc}$ is the mass of the stellar galactic
nucleus in the dwarf galaxy for the standard dwarf model (D1).
\label{fig-12}}
\end{figure}

\end{document}